\documentclass [11pt]{article}
\usepackage{amsfonts}
\usepackage{graphicx}
\usepackage{amsmath,amssymb, amsfonts}
\usepackage{latexsym}
\usepackage{mathrsfs}
\usepackage{bbold}
\usepackage{color}
\usepackage{slashed}
\usepackage{cancel}
\usepackage{soul} %
\usepackage{setspace} %
\usepackage{comment}
\usepackage{booktabs}
\usepackage{multirow}
\usepackage{tikz-feynman}
\tikzfeynmanset{compat=1.0.0}
\usepackage{cite}
\usepackage[small]{caption}
\usepackage{epsfig}
\usepackage{slashed}
\usepackage{epstopdf}
\usepackage{latexsym}
\usepackage{graphicx}
\usepackage{subcaption}
\usepackage{tikz}
\usetikzlibrary{decorations.pathmorphing}
\pgfdeclaredecoration{complete sines}{initial}
{
    \state{initial}[
        width=+0pt,
        next state=sine,
        persistent precomputation={\pgfmathsetmacro\matchinglength{
            \pgfdecoratedinputsegmentlength / int(\pgfdecoratedinputsegmentlength/\pgfdecorationsegmentlength)}
            \setlength{\pgfdecorationsegmentlength}{\matchinglength pt}
        }] {}
    \state{sine}[width=\pgfdecorationsegmentlength]{
        \pgfpathsine{\pgfpoint{0.25\pgfdecorationsegmentlength}{0.5\pgfdecorationsegmentamplitude}}
        \pgfpathcosine{\pgfpoint{0.25\pgfdecorationsegmentlength}{-0.5\pgfdecorationsegmentamplitude}}
        \pgfpathsine{\pgfpoint{0.25\pgfdecorationsegmentlength}{-0.5\pgfdecorationsegmentamplitude}}
        \pgfpathcosine{\pgfpoint{0.25\pgfdecorationsegmentlength}{0.5\pgfdecorationsegmentamplitude}}
}
    \state{final}{}
}

\usepackage{rotating}
\usepackage{bbm}
\usepackage{bbold}
\usepackage{enumitem}

\usepackage{empheq}

\definecolor{myblue}{rgb}{.8, .8, 1}
\newcommand*\mybluebox[1]{%
\colorbox{myblue}{\hspace{0.5em}#1\hspace{0.5em}}}

\definecolor{red}{rgb}{1,0,0}
\usetikzlibrary{positioning}
\tikzset{
    mybluenode/.style={
        draw=black, circle, minimum width=2cm, inner sep=0pt
        },
    myblacknode/.style={
        circle, inner sep=1pt, fill=black
        },
    }

\newcommand*\be{\beta}

\newcommand*\G{\Gamma}
\newcommand*\di{\delta}

\newcommand*\la{\lambda}

\newcommand*\Di{\Delta}

\newcommand*\RR{\mathbb{R}}

\newcommand*\calO{\mathcal{O}}
\newcommand*\calP{\mathcal{P}}

\newcommand*\intads{\int}
\makeatletter
\def\section{\@startsection {section}{1}{\z@}{-3.5ex plus -1ex minus
 -.2ex}{2.3ex plus .2ex}{\large\bf}}
\def\subsection{\@startsection{subsection}{2}{\z@}{-3.25ex plus -1ex
minus -.2ex}{1.5ex plus .2ex}{\normalsize\bf}}
\makeatother
\makeatletter

\@addtoreset{equation}{section}

\makeatother

\newcommand{\bea}{\begin{equation} \begin{aligned}} \newcommand{\eea}{\end{aligned} \end{equation}}
\def\be{\begin{equation}} \def\ee{\end{equation}} 
\def\nn{\nonumber}

\usepackage[colorlinks,linkcolor=black,citecolor=blue,urlcolor=blue,linktocpage,pagebackref]{hyperref}

\allowdisplaybreaks

\begin{document}

\thispagestyle{empty}

\begin{center}

	\vspace*{-.6cm}

	\begin{center}

		\vspace*{1.1cm}

		{\centering \Large\textbf{Exploring Confinement in Anti-de Sitter Space}}

	\end{center}

	\vspace{0.8cm}
	{\bf Riccardo Ciccone$^{d,e}$, Fabiana De Cesare$^{b,c}$, Lorenzo Di Pietro$^{a,b}$, Marco Serone$^{b,c}$}

	\vspace{1.cm}
	
	${}^a\!\!$
	{\em  Dipartimento di Fisica, Universit\`a di Trieste, \\ Strada Costiera 11, I-34151 Trieste, Italy}
		
	\vspace{.3cm}

	${}^b\!\!$
	{\em INFN, Sezione di Trieste, Via Valerio 2, I-34127 Trieste, Italy}

	\vspace{.3cm}

	${}^c\!\!$
	{\em SISSA, Via Bonomea 265, I-34136 Trieste, Italy}

	\vspace{.3cm}

 ${}^d\!\!$
	{\em Department of Physics and Haifa Research Center for Theoretical\\
Physics and Astrophysics, University of Haifa, Haifa 31905, Israel}

	\vspace{.3cm}
 ${}^e\!\!$
	{\em Department of Physics, Technion,\\ Israel Institute of Technology, 32000 Haifa, Israel}

	\vspace{.3cm}

\end{center}

\vspace{1cm}

\centerline{\bf Abstract}
\vspace{2 mm}
\begin{quote}

We study Yang-Mills theory on four dimensional Anti-de Sitter space. The Dirichlet boundary condition cannot exist at arbitrarily large radius because it would give rise to colored asymptotic states in flat space. As observed in \cite{Aharony:2012jf} this implies a deconfinement-confinement transition as the radius is increased. We gather hints on the nature of this transition using perturbation theory. We compute the anomalous dimensions of the lightest scalar operators in the boundary theory, finding that the singlet gets a larger negative anomalous dimension compared to the adjoint.
We also compute the correction to the coefficient $C_J$ and we estimate that the singlet operator reaches marginality before the value of the coupling at which $C_J=0$. These results favor the scenario of merger and annihilation as the most promising candidate for the transition. For the Neumann boundary condition, the lightest scalar operator is found to have a positive anomalous dimension, in agreement with the idea that this boundary condition extrapolates smoothly to flat space. The perturbative calculations are made possible by a drastic simplification of the gauge field propagator in Fried-Yennie gauge. We also derive a general result for the leading-order anomalous dimension of the displacement operator for a generic perturbation in Anti-de Sitter, showing that it is related to the beta function of bulk couplings.

\end{quote}

\newpage

\tableofcontents
	
\newpage 
\section{Introduction}

Establishing the existence of the mass gap in Yang-Mills (YM) theory is a major open problem in theoretical and mathematical physics \cite{Jaffe:2000ne,Doug2004}. A related, even stronger, condition that is believed to hold is color confinement, namely the property that all the asymptotic states are massive color-singlet particles. Both of these properties pertain to the large-distance behavior of observables, making it difficult to reach quantitative control due to the strong coupling of the interactions.

A possible approach to try to make progress on these questions is to cut off the large IR fluctuations, by putting the theory in a box of size $L$ and then studying the limit of $L\to\infty$. This can be done without sacrificing spacetime symmetries if the ``box'' is a maximally symmetric curved space of radius $L$.  Among those spaces, Anti-de Sitter (AdS) space, thanks to the existence of a conformal boundary, has the advantage of admitting a clear notion analogous to that of asymptotic states in flat space, namely the states associated with the insertions of local operators at the boundary. Connecting these states to scattering states in the flat-space limit is a well-studied problem \cite{Polchinski:1999ry,Giddings:1999qu,Fitzpatrick:2010zm,Penedones:2010ue,Paulos:2016fap,Mazac:2016qev,Mazac:2018mdx,Komatsu:2020sag,Li:2021snj,Knop:2022viy,Cordova:2022pbl,Gadde:2022ghy,vanRees:2022zmr,vanRees:2023fcf}. The isometries of the background ensure that the correlators of the boundary operators encompass a conformal field theory (CFT), whose operator content and data depend on the choice of boundary conditions for the bulk fields.\footnote{In this paper we take AdS to be a rigid background and we do not include dynamical gravity. As a result the CFT on the boundary does not have a stress tensor operator.} Moreover, AdS space has infinite volume even at finite $L$, allowing the possibility of spontaneous symmetry breaking and phase transitions, phenomena which are forbidden on compact spaces. Thanks to these nice properties, quantum field theory in rigid AdS has been the object of a revived interest, see e.g.  \cite{Carmi:2018qzm, Hogervorst:2021spa, Giombi:2020rmc, Giombi:2021cnr, Antunes:2021abs, Ankur:2023lum, Lauria:2023uca, Meineri:2023mps, Copetti:2023sya, Levine:2023ywq, Antunes:2024hrt} for a partial list of recent developments. 

The study of four-dimensional non-abelian gauge theories on the background of Euclidean AdS space, i.e. hyperbolic space, was advocated long ago in \cite{Callan:1989em} as a way to have better control on the non-perturbative effects. The meaning of confinement in AdS space was later explored in \cite{Aharony:2012jf}, which pointed out the existence of a deconfinement-confinement transition as the radius $L$ is increased. When $L$ is small in units of the dynamically-generated scale $\Lambda_{\text{YM}}$ the theory can be placed in AdS by imposing the standard Dirichlet boundary condition ($D$ bc) for the gauge fields,
\begin{equation}
A^a_i(x,z) \underset{z\to 0}{\sim} z \,g^2\,J^a_i(x)\,,
\end{equation}
and the bulk gauge symmetry $G$ becomes a global symmetry on the boundary. We are now restricting to (Euclidean) AdS$_4$, with metric
\begin{equation}
ds^2 = L^2 \frac{d z^2 + dx_i^2}{z^2}\,,~~i=1,2,3\,,
\end{equation}
and $g^2$ is the YM coupling. The spectrum of operators in the boundary CFT
contains conserved currents $J^a_i$ of the non-abelian symmetry $G$, and more generally operators in non-trivial representations of $G$. Moreover, the conserved currents cannot continuously recombine at $g^2=0$ and therefore they keep the protected dimension $\Delta_J=2$ for a finite range of $L$ around $L=0$. If these operators would still exist for arbitrarily large $L$, they would give rise to asymptotic states of massless gluons in the limit $L\to\infty$. Therefore \cite{Aharony:2012jf} argued that a necessary condition for the existence of the mass gap in flat space is that there is a transition at some finite value of $L$ to a different boundary condition, one in which the currents and the associated symmetry are not present on the boundary.
This has to be contrasted with the case of the Neumann boundary condition ($N$ bc),
\begin{equation}
A^a_i(x,z) \underset{z\to 0}{\sim}  a^a_i(x)\,,
\end{equation}
where the boundary mode is a 3d gauge field, and the group $G$ remains a gauge symmetry of the boundary. As a result in this case all the physical operators of the boundary CFT are color singlets, and it is possible for this boundary condition to smoothly approach the flat-space limit $L\to \infty$. 

The deconfinement-confinement transition is expected to happen at strong coupling, a natural estimate for the critical radius being $L_{\text{crit}}\sim \Lambda_{\text{YM}}^{-1}$, and therefore it is hard to make precise statements about it. Various alternative mechanisms for the transition can be envisioned, as explained in \cite{Aharony:2012jf}
and recently revisited in \cite{Copetti:2023sya}:
\begin{itemize}
\item{{\bf Higgsing:} A scalar operator $O^a$ in the adjoint representation of $G$ becomes marginal at $L_\text{crit}$ and recombines with the current $\partial^i J_i^a = O^a$, allowing the latter to get an anomalous dimension and breaking the $G$ global symmetry;}
\item{{\bf Decoupling:} The positive coefficient $C_J$ of the current two-point function, which gives the norm of the state associated with the current operator, goes to $0$ at $L_\text{crit}$, forcing the current operator to decouple from the theory;} 
\item{{\bf Marginality:} A singlet scalar operator $O$ becomes marginal at $L_\text{crit}$, causing the $D$ bc to merge and annihilate \cite{Kaplan:2009kr, Gorbenko:2018ncu} with a second one, and to stop existing as a unitary boundary condition.}
\end{itemize}
The third mechanism was advocated as the most likely in \cite{Copetti:2023sya}, based on the analogy with 2d asymptotically-free models which can be studied in the $1/N$ expansion.

Proving the existence of the transition in which the $D$ bc disappears is a very interesting problem. Understanding the nature of the transition and having quantitative control over it could potentially offer new perspectives on the mass gap and confinement problem. In this paper, we investigate this problem using perturbation theory. We will argue that perturbation theory can play a valuable role in discerning between the various proposed scenarios, besides providing data that can be later used as inputs for the numerical conformal bootstrap. Working in an expansion around small radius, or equivalently in the Yang-Mills coupling $g^2$ at the scale $L^{-1}$, we compute the following quantities at next-to-leading order (NLO):
\begin{itemize}
\item{the scaling dimension of the lightest singlet scalar operator, both for $D$ and $N$ bc's;}
\item{the scaling dimensions of the lightest scalar operators in non-trivial representations of the $G$ global symmetry for the $D$ bc;}
\item{the coefficient $C_J$ of the current two-point function for the $D$ bc.}
\end{itemize}
These are the quantities that are more directly related to the possible scenarios for the transition. Moreover, the quantity $C_J$ is the CFT proxy for the bulk gauge coupling, and our result allows to map any bulk calculation in an expansion in the gauge coupling in dimensional regularization to an expansion in $1/C_J$, up to NLO. 

\paragraph{Results} Considering for definiteness $G=SU(n_c)$, we find that
the lightest singlet scalar operator in the $D$ bc, namely $\mathrm{tr}[J_i J^i]$, has the following negative anomalous dimension
\begin{equation}\label{eq:dimJJ}
D: \quad \Delta_{\mathrm{tr}[JJ]} = 4 - \frac{11 n_c}{24\pi^2} \,g^2 +\mathcal{O}(g^4)\,.
\end{equation}
The lightest adjoint scalar, which is also bilinear in the currents at weak coupling and therefore also starts from dimension $4$, gets a negative anomalous dimension smaller in absolute value by a factor of 2, see \eqref{eq:gammaJJadjoint}.\footnote{{This applies for $n_c>2$. When $n_c=2$ the Higgsing scenario is strongly disfavoured, as the lightest adjoint scalar has classical dimension $\Delta=7$.}} This indicates that the Marginality scenario is more likely than the Higgsing one, in agreement with  \cite{Copetti:2023sya}. Truncating at NLO we can roughly estimate the transition to happen at 
\begin{equation}\label{eq:estimate}
\Delta_{\mathrm{tr}[JJ]} = 3 ~~\Rightarrow~~ g^2_{\text{crit}}\vert_{\text{NLO}} = \frac{24\pi^2}{11 n_c} \approx \frac{21.5}{n_c}\,,~\text{ or equivalently}~(L\Lambda_{\text{YM}})_{\text{crit}}\vert_{\text{NLO}} = \frac{1}{e}\approx 0.37\,. 
\end{equation}
This estimate needs to be taken with caution because it is just the result of a one-loop truncation. Nevertheless, we note that the estimated value of $n_cg^2_{\text{crit}}/(16\pi^2)$ is quite small, suggesting that perturbation theory might still be sufficiently reliable, in contrast to the expectation that the transition should happen at strong coupling.\footnote{For comparison, Higgsing {for $n_c>2$} is estimated to occur at $(L\Lambda_{\text{YM}})_{\text{crit}}\vert_{\text{NLO}} =e^{-1/2}\approx 0.61$.} The indication towards Marginality is further confirmed by the fact that for the $N$ bc instead the lightest scalar singlet operator, namely $\mathrm{tr}[f_{ij}f^{ij}]$, has positive anomalous dimension,
\begin{equation}\label{eq:dimff}
N: \quad \Delta_{\mathrm{tr}[ff]} = 4 + \frac{11 n_c}{24\pi^2} \,g^2 +\mathcal{O}(g^4)\,.
\end{equation}
This agrees with the idea that the $N$ bc smoothly interpolates to the flat space limit, and therefore no singlet operator is expected to cross marginality. The first correction to $C_J$ also happens to be negative,
\begin{equation}
C_J = \frac{2}{\pi^2 g^2}\left(1-\frac{10 + 3 \gamma_E}{324 \pi^2}\ n_c\, g^2 +\mathcal{O}(g^4)\right)\,,
\end{equation}
but the NLO estimate of the critical coupling in the Decoupling scenario gives 
\begin{equation}\label{eq:Decoup}
C_J= 0 ~~\Rightarrow~~ g^2_{\text{crit}}\vert_{\text{NLO}}   \approx \frac{272}{n_c}\,,~\text{ or equivalently}~(L\Lambda_{\text{YM}})_{\text{crit}}\vert_{\text{NLO}} \approx 0.92\,. 
\end{equation}
Compared to \eqref{eq:estimate}, this estimate suggests that the transition in the Marginality scenario happens before.

\begin{figure}
\centering

\tikzset{every picture/.style={line width=0.75pt}} 

\begin{tikzpicture}[x=0.75pt,y=0.75pt,yscale=-1,xscale=1]

\draw [line width=1.5]    (100,129) -- (565.5,129.99) ;
\draw [shift={(568.5,130)}, rotate = 180.12] [color={rgb, 255:red, 0; green, 0; blue, 0 }  ][line width=1.5]    (19.89,-5.99) .. controls (12.65,-2.54) and (6.02,-0.55) .. (0,0) .. controls (6.02,0.55) and (12.65,2.54) .. (19.89,5.99)   ;
\draw [color={rgb, 255:red, 126; green, 211; blue, 33 }  ,draw opacity=1 ][line width=1.5]    (99.5,11) .. controls (193.5,12) and (340.5,5) .. (340.5,39) ;
\draw [color={rgb, 255:red, 208; green, 2; blue, 27 }  ,draw opacity=1 ][line width=1.5]    (236.5,67) .. controls (282.5,67) and (340.5,61) .. (340.5,41) ;
\draw [color={rgb, 255:red, 208; green, 2; blue, 27 }  ,draw opacity=1 ][line width=1.5]  [dash pattern={on 5.63pt off 4.5pt}]  (180.5,67) .. controls (194.5,67) and (212.5,67) .. (230.5,67) ;
\draw  [fill={rgb, 255:red, 0; green, 0; blue, 0 }  ,fill opacity=1 ] (335.13,37.62) .. controls (335.89,35.21) and (338.47,33.87) .. (340.88,34.63) .. controls (343.29,35.39) and (344.63,37.97) .. (343.87,40.38) .. controls (343.11,42.79) and (340.53,44.13) .. (338.12,43.37) .. controls (335.71,42.61) and (334.37,40.03) .. (335.13,37.62) -- cycle ;
\draw [color={rgb, 255:red, 74; green, 144; blue, 226 }  ,draw opacity=1 ][line width=1.5]    (100.5,98) -- (565.5,98) ;
\draw  [dash pattern={on 4.5pt off 4.5pt}]  (340.5,49) -- (340.5,130) ;

\draw (542,140) node [anchor=north west][inner sep=0.75pt]   [align=left] {$\displaystyle L \Lambda_{\text{YM}}$};
\draw (319,137) node [anchor=north west][inner sep=0.75pt]   [align=left] {$\displaystyle ( L \Lambda_{\text{YM}})_{\text{crit}}$};
\draw (125,18) node [anchor=north west][inner sep=0.75pt]  [color={rgb, 255:red, 126; green, 211; blue, 33 }  ,opacity=1 ] [align=left] {$\displaystyle D$};
\draw (243.5,71) node [anchor=north west][inner sep=0.75pt]  [color={rgb, 255:red, 208; green, 2; blue, 27 }  ,opacity=1 ] [align=left] {$\displaystyle D^{*}$};
\draw (126,103) node [anchor=north west][inner sep=0.75pt]  [color={rgb, 255:red, 74; green, 144; blue, 226 }  ,opacity=1 ] [align=left] {$\displaystyle N$};
\draw (164,59) node [anchor=north west][inner sep=0.75pt]   [align=left] {?};

\end{tikzpicture}
\caption{Schematic representation of the conjectured evolution of the $D$ and $N$ bc's as a function of the bulk coupling $L\Lambda_{\text{YM}}$. The $D$ bc merges and annihilates with $D^*$, a second boundary condition with $G$ global symmetry, which must exist for $L\Lambda_{\text{YM}} \lesssim (L\Lambda_{\text{YM}})_{\text{crit}}$, but is not guaranteed to exist at weak coupling. \label{fig:DDstarN}}
\end{figure}
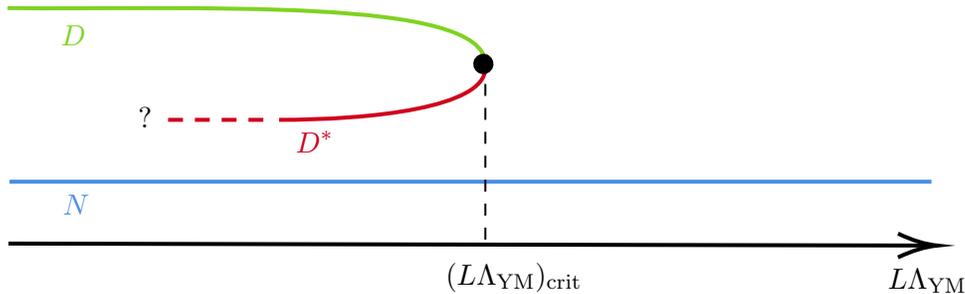

The conjectural picture that is suggested by these results is illustrated in figure \ref{fig:DDstarN}. The $D$ bc exists for a finite range of $L\Lambda_{\text{YM}}$, but at the critical value $(L\Lambda_{\text{YM}})_\text{crit}$
the operator $\mathrm{tr}[JJ]$ becomes marginal and the associated boundary coupling $\eta$ has a beta function \cite{Lauria:2023uca}
\begin{equation}\label{eq:betamerging}
\beta_\eta = c_1 \eta^2 + c_2 \left(\frac{1}{g^2}-\frac{1}{g^2_\text{crit}}\right)\,\quad \text{for} \quad g^2\approx g^2_\text{crit}\,, \;\; \eta\ll 1 \,.
\end{equation}
As we review, the coefficients $c_{1,2}$ can be expressed in terms of data of the boundary CFT for $g^2= g^2_\text{crit}$, whose value is not calculable. Even without knowing their values,
the existence of the $D$ bc for $g^2 < g^2_{\text{crit}}$ ensures that the condition $\beta_\eta = 0$ must have two real solutions for $g^2 \lesssim g^2_\text{crit}$, one of them being the $D$ bc, and the second being an additional boundary condition with $G$ global symmetry, which we call $D^*$.  
The theories $D$ and $D^*$ merge and annihilate at $g^2_\text{crit}$, and become complex at larger values of the coupling. In this way, the $D$ bc stops being 
a viable boundary condition for $g^2>g^2_{\text{crit}}$. This picture raises the question of better understanding the nature of the $D^*$ bc, which we will not study in this paper. On the other hand, the $N$ bc is suggested to exist for all values of $(L\Lambda_{\text{YM}})$, as envisioned in \cite{Aharony:2012jf}.

Besides the standard perturbative computation, we discuss a different approach for the calculation of the anomalous dimensions of $\mathrm{tr}[JJ]$ in the $D$ bc and $\mathrm{tr}[ff]$ in the $N$ bc. This approach is based on the fact that in the limit $g^2\to 0$ the bulk theory is the free UV CFT of YM theory. As a result, the whole setup can be mapped via a Weyl rescaling to flat-space boundary CFT (BCFT), and the two operators can be identified with the so-called displacement operator of the respective boundary condition. We provide a general argument based on multiplet recombination that fixes the anomalous dimension of the displacement operator for a generic perturbation of a CFT in AdS, see \eqref{eq:anodimDgen}. In particular, we find that for a classically marginal deformation the leading anomalous dimension is determined by the one-loop beta function in the bulk, see \eqref{eq:anodimDmarg}. This is the reason why the coefficient of the one-loop beta function of YM theory appears in \eqref{eq:dimJJ} and \eqref{eq:dimff}. The result in the case of the $D$ bc is then matched with the explicit diagrammatic calculation.

In preparation for the perturbative calculation, we also discuss in detail the propagators for gauge fields in AdS$_{d+1}$ (generic $d$ is needed for dimensional regularization) with $R_\xi$ gauge fixing, both in $D$ and $N$ bc's. We find that choosing the gauge fixing parameter as $\xi = \frac{d}{d-2}$, i.e. the Fried-Yennie (FY) gauge \cite{Fried:1958zz, Yennie:1961ad}, leads to drastic simplifications in the propagators, e.g. from derivatives of hypergeometric functions to rational functions. For instance, for the $D$ bc the expression \eqref{eq:bulkDanyxi} for the bulk-to-bulk propagators in generic $\xi$ collapses to \eqref{subYennie} in FY gauge. It is only thanks to these remarkable simplifications that we are able to carry through the brute-force calculation of the diagrams in position space.

\paragraph{Outline} The rest of the paper is organized as follows: in section \ref{sec:ads} we discuss the generalities of YM in AdS space and the $D$ and $N$ bc's at weak coupling, in particular the spectrum of boundary operators in the free limit; in section \ref{sec:adsprop} we derive the propagators for gauge fields in AdS in $R_\xi$ gauge, and the special properties of the FY gauge; in section \ref{sec:gammaJJ} we first present the general multiplet recombination argument for the anomalous dimension of the displacement operator induced by AdS deformations, then we apply it to compute the anomalous dimensions of the lightest singlet operators for the $D$ and $N$ bc's, and finally we perform the diagrammatic calculation of the anomalous dimensions of all current bilinear operators in the $D$ bc, both singlet and non-singlets; in section \ref{sec:cj} we perform the perturbative calculation of $C_J$; in section \ref{sec:conc} we conclude and discuss possible future directions. Several appendices contain technical details, and a review of the calculation of beta functions for boundary marginal operators.

\paragraph{Conventions} Throughout the paper we use late lowercase Greek letters $\mu,\nu,\ldots$ for indices on Euclidean AdS$_{d+1}$ space, early lowercase Latin letters $a,b,\ldots$ for gauge group indices, late lowercase Latin letters $i,j,\ldots$ for indices on $\mathbb{R}^d$, and early uppercase Latin letters $A,B,\ldots$ for embedding space indices. We use the following notation for integration over bulk 
points,
\begin{equation}\label{eq:intconv}
    \int dx \,f(x)\equiv \int_{{\rm AdS}} d^{d+1}x\sqrt{g(x)}\ f(x)\,,
\end{equation}
where $g(x)$ is the determinant of the AdS metric $g_{\mu\nu}$. In embedding coordinates, the integral \eqref{eq:intconv} is expressed as
\begin{equation}
\begin{aligned}
    \int dX\,f(X)&\equiv \int d^{d+2}X\ \delta(X^2+L^2)\Theta(X^0)f(X)\,,
    \end{aligned}
\end{equation}
where $\Theta(x)$ is the Heaviside step function. 

\section{Generalities of YM theory on AdS and boundary theory}
\label{sec:ads}

The action of Yang-Mills theory on Euclidean AdS$_{d+1}$ in $\xi$-gauge reads (ghost terms omitted)
\begin{equation}
   S_\text{YM}= \frac{1}{g^2} \int dx\   \mathrm{tr} \bigg[\frac 12 F_{\mu \nu} F^{\mu \nu} +  \frac{1}{\xi} (\nabla_\mu A^\mu)^2\bigg],
   \label{eq:gpAdS1}
\end{equation}
where 
$F_{\mu \nu} = F_{\mu \nu}^a t_a$, with
\begin{equation}
    F_{\mu \nu}^a = \partial_\mu A_\nu^a - \partial_\nu A_\mu^a + f^{a}{}_{bc} A_\mu^b A_\nu^c\,,
    \label{eq:gpAdS2}
\end{equation}
and {${\rm tr} [t_a t_b] = \delta_{ab}/2$}.
We are interested in the physics for $d=3$, but keeping $d$ generic is needed for dimensional regularization.  We will mostly use Poincaré coordinates $x^\mu=(x^i,z)$ with $z> 0$ and $i=1,\dots,d$, in which the metric $g_{\mu \nu}$ reads 
\begin{equation}\label{eq:MPoin}
d s^2=g_{\mu\nu} dx^\mu dx^\nu =L^2 \frac{d z^2+dx_i^{2}}{z^2}\,.
\end{equation}
$L$ is the radius of AdS, which we set to 1  unless explicitly specified. The boundary is at $z=0$. 

The allowed boundary conditions of $A_\mu$ can be worked out by looking at the behavior of the equations of motion close to $z= 0$. One has \cite{Witten:2003ya,Marolf:2006nd}
\begin{equation}\label{eq:BoundLimDN}
\begin{split}
D:~~A_i^a(x,z) & \underset{z\to 0}{\sim} z^{d-2} g^2 J_i^a(x)\,,  \\
N:~~A_i^a(x,z) & \underset{z\to 0}{\sim}  a_i^a(x)\,. 
\end{split}
\end{equation}
In the first case, for $d>2$ the bulk gauge field vanishes at the boundary and we have $D$ bc, with $J_i^a(x)$ a conserved non-abelian vector current with scaling dimension $\Delta_J = d-1$.
In the second case, the bulk gauge field does not vanish at the boundary and we have $N$ bc, with $a_i^a(x)$ a non-abelian gauge connection with scaling
dimension $\Delta_a = 1$. This value is below the unitarity bound for spin 1 operators but this is not an issue because $a_i^a$ is not a gauge-invariant primary operator. 

We will mostly focus on $D$ bc. When the gauge interactions are switched off, the bulk gauge field $A$ gives rise to a boundary CFT which is the mean-field theory \cite{Heemskerk:2009pn}
of the non-abelian conserved currents $J_i^a$. Their two-point function read 
 \begin{equation} \label{eq:JJ2pt}
    \langle  J^{a}_{i}(x_1) J^{b}_{j}(x_2)\rangle =\frac{ C_J^0}{g^2} \delta^{a b} \frac{I_{ij}}{x_{12}^{2(d-1)}}\,,
    \end{equation}
where $x_{12} = x_1-x_2$, and
\begin{equation}
    I_{ij} =   \delta_{i j}-\frac{2(x_{12})_i (x_{12})_j}{x_{12}^2}~,~~ C_J^0 = \frac{\Gamma(d)}{ 2(d-2)\pi^{\frac{d}{2}}\Gamma(\tfrac d2)}\,.
    \label{eq:CJ0}
\end{equation}
Let us now focus on the case of $d=3$. At $g^2=0$ the $D$ boundary theory is given by all the primary operators of the schematic form 
$J^{n_1}\Box^p \partial^m J^{n_2}$ with correlation functions entirely determined by \eqref{eq:JJ2pt} and Wick's contractions. In table \ref{tab:doubletraceprimaries} we report the first ``double trace" $JJ$ primary operators up to $\Delta \leq 7$, including their representation under the global symmetry, 
taken to be $G=SU(n_c)$ for definiteness. This is obtained using the standard technique based on 
characters and the partition function of single particle states \cite{Sundborg:1999ue,Aharony:2003sx}. The non-abelian structure allows for more primaries than those appearing in the mean-field theory of an abelian $U(1)$ current, as Bose symmetrization of $J_i^a$, $J_j^b$ can be achieved either by symmetrizing or anti-symmetrizing both their spacetime and adjoint flavor indices. When restricted to the singlet flavor representation, the spectrum of double trace primary operators in table \ref{tab:doubletraceprimaries} reduces to that obtained from an abelian current $J_i$, see e.g. \cite{Karateev:2018oml}. 
We refer the reader to appendix \ref{app:characters} for details and for the complete list of operators up to $\Delta \leq 7$ which includes ``triple trace" operators. When $g^2\neq 0$, the interactions mix operators with the same quantum numbers, which also get anomalous dimensions.

\begin{table}[]
\renewcommand{\arraystretch}{1.15}
\centering
\begin{tabular}{|c|c|c|}
\hline
$\calO$                                             & $(\Delta,\ell)_\pi$ & $R(SU(n_c))$             \\ \hline
$[J_i^a J_i^b]$                                   & (4,0)$_+$           & $R_+$           \\
$[J_{i}^a J_{j}^b]$                           & (4,2)$_+$           & $R_+$           \\
$[J_i^a \partial_i J_j^b]$                           & (5,1)$_+$           & $R_-$           \\
$[J_{i}^a \partial_j J_{k}^b]$             & (5,3)$_+$           & $R_-$   \\
$[J_{i}^a \square J_{i}^b]$             & (6,0)$_+$           & $R_+$   \\
$[J_i^a \square J_j^b]$ & (6,2)$_+$           & $R_+$           \\
$[\widetilde{J}^a_i\widetilde{J}^b_j]$& (6,2)$_+$           & $R_+$           \\
$[J_{i}^a \partial_j\partial_k J_{l}^b]$                           & (6,4)$_+$           & $R_+$           \\
$[J_i^a \square\partial_i J_j^b]$                           & (7,1)$_+$           & $R_-$           \\
$[\widetilde{J}_{i}^a \partial_{j} \widetilde{J}_{k}^b]$             & (7,3)$_+$           & $R_-$   \\
$[J_{i}^a \square\partial_j J_{k}^b]$             & (7,3)$_+$           & $R_-$   \\
$[J_{i}^a \partial_j\partial_k\partial_l J_{m}^b]$             & (7,5)$_+$           & $R_-$   \\

\hline       
\end{tabular} \hspace{1cm}
\begin{tabular}{|c|c|c|}
\hline
$\calO$                                             & $(\Delta,\ell)_\pi$ & $R(SU(n_c))$             \\ \hline

$[\epsilon_{ijk}J_{i}^aJ_{j}^b]$           & (4,1)$_-$           & $R_-$           \\
$[J_i^a\widetilde{J}_i^b]$
& (5,0)$_-$           & $R_+$           \\
$[J_i^a\widetilde{J}_j^b]$ & (5,2)$_-$           & $R_+\oplus R_-$ \\
$[J_{i}^a \partial_j \widetilde{J}_{i}^b]$
& (6,1)$_-$           & $R_-$   \\
$[J_{i}^a \partial_j \widetilde{J}_{k}^b]$ & (6,3)$_-$           & $R_+\oplus R_-$           \\
$[J_i^a\square \widetilde{J}_i^b]$
& (7,0)$_-$           & $R_+$           \\
$[J_i^a\square \widetilde{J}_j^b]$& (7,2)$_-$           & $R_+\oplus R_-$ \\
$[J_i^a\partial_j\partial_k \widetilde{J}_l^b]$& (7,4)$_-$           & $R_+\oplus R_-$   \\
\hline       
\end{tabular}
\caption{Double trace $JJ$ primary operators up to $\Delta\leq7$ in the mean field theory of $SU(n_c)$ adjoint currents, $\calO$ being their schematic form. We have distinguished between parity-even $\pi=+1$ primaries (left) and parity-odd $\pi=-1$ primaries (right). Here $\widetilde{J}_i^a\equiv\epsilon_{ijk}\partial_j J^a_k$. $R_+$ (resp. $R_-$) labels the $SU(n_c)$ representations corresponding to the symmetric (resp. antisymmetric) product of two adjoint representations, see appendix \ref{app:characters} for details. In particular, $R_+$ always contains the singlet $\mathbf{1}$ and the adjoint representation $R_A$.}
\label{tab:doubletraceprimaries}
\end{table}

At $g^2=0$ the $N$ boundary theory is the direct sum of $n_c^2-1$ mean field theories of abelian antisymmetric tensor fields, $f_{ij}^a$,
at the unitarity bound.
These fields are dual to the currents: $J_i^a = \epsilon_{ijk} f_{jk}^a$, which are conserved due to the Bianchi identity satisfied by $f_{ij}^a$.
The table \ref{tab:doubletraceprimaries} can then be used also to extract the spectrum of local operators of the $N$ CFT at $g^2=0$. However, these results have to be interpreted with care when interactions are turned on. At $g^2\neq 0$, all non-singlet operators become unphysical and we are left only with the singlet ones, which mix among each other and would in general acquire anomalous dimensions.\footnote{The presence of many more local operators in the $g^2=0$ theory can be seen to arise from endpoints of line operators, which in the limit $g^2\rightarrow 0$ become local operators.}

\section{The gauge propagator in AdS}
\label{sec:adsprop}

In spaces with boundaries, we can distinguish between bulk-to-bulk, bulk-to-boundary, and boundary-to-boundary propagators.
The knowledge of the former clearly allows to derive the other two by sending the bulk points to the boundary. 
Bulk-to-boundary gauge propagators in AdS (with $D$ bc) are entirely fixed by conformal symmetry, are $\xi$-independent, and have been determined since the first years of the AdS/CFT correspondence \cite{Witten:1998qj,Freedman:1998tz}. In contrast, bulk-to-bulk gauge propagators are significantly more involved. 
Bulk-to-bulk propagators for massive spin 1 fields have been determined in ambient space in \cite{Naqvi:1999va,Leonhardt:2003sn} and rederived more elegantly using embedding
space techniques in \cite{Costa:2014kfa}. As far as we know the only computation of the bulk-to-bulk gauge propagator for massless gauge fields in configuration space dates back to \cite{Allen1986}, where it has been computed in the Feynman gauge $\xi=1$ using ambient space techniques. The resulting expression is quite complicated and consists of a sum of hypergeometric functions and their derivatives with respect to the parameters $a,b,c$.\footnote{An expression for the bulk-to-bulk gauge propagator, that however neglects matters related to gauge-fixing, was given in \cite{DHoker:1998bqu,DHoker:1999bve}. 
Moreover, very recently the bulk-to-bulk gauge propagator has been determined in a mixed momentum-configuration space in the $A_z=0$ and the Landau $\xi=0$ gauges \cite{Marotta:2024sce}.} 

In this section, we compute the bulk-to-bulk gauge propagator in AdS$_{d+1}$ space by using techniques of harmonic analysis, for any $\xi$-gauge. Quite remarkably, we find
that the propagator dramatically simplifies for an appropriate gauge choice.

It is useful to adopt embedding coordinates to write the propagators. Embedding space techniques for AdS${}_{d+1}$ have been worked out in \cite{Costa:2014kfa}, building on previous work where they have been developed for CFTs in $d$ dimensions \cite{Costa:2011mg}. 
Euclidean AdS$_{d+1}$ can be embedded into $(d+2)-$dimensional Minkowski space $\mathbb{R}^{d+1,1}$ with coordinates $X^A$ as
\begin{equation} 
X^2=\eta_{A B} X^A X^B=-{1}\,,
\label{eq:gpAdS4}
\end{equation}
where $\eta_{AB}={\rm diag}\,(-+\dots+)$. Tensors in embedding space can be encoded in a convenient index-free notation. We use here the conventions of \cite{Costa:2014kfa} and refer the reader to this reference for explanations. 
Expressions in the AdS$_{d+1}$ ambient space are given in the Poincar\'e coordinates $x^\mu \equiv (x^i,z)$,  which are related as follows to embedding coordinates,
\begin{equation}
    X=(X^+,X^-,X^i)=\frac 1z (1,x^2+z^2,x^i)\,,
\end{equation}
where $x^2 = x_i x^i$, and $X^\pm={X^{0}\pm X^{d+1}}$ are light-cone coordinates on $\mathbb{R}^{d+1,1}$.
We parametrize the distance between points $x^\mu =\left(x^i,z\right)$ and $y^\mu =\left(y^i,w\right)$ by
\begin{equation}
u(x,y)=\frac{\left(x-y\right)^2+(z-w)^2}{2 z w} =\frac{(X-Y)^2}{2}=-(1+ X \cdot Y) \,.
\end{equation}

The gluon propagator $\langle A_\mu (x)A_\nu (y)\rangle:=g^2\Pi_{\mu\nu}$ satisfies the following equation in ambient space,
\begin{equation}
   \left(-\delta^\rho_\mu\nabla^2+R^\rho_\mu+\left(1-\tfrac{1}{\xi}\right)\nabla^\rho\nabla_\mu\right) \Pi_{\rho\nu}(x,y) = g_{\mu\nu}\delta(x-y) \,,
 \label{eq:gpAdS3}
\end{equation}
where $R_{\mu\nu}= - d  g_{\mu \nu}$. The uplift in embedding space of \eqref{eq:gpAdS3}, in index-free notation, reads
\begin{equation}\label{eq:spin1P}
\begin{aligned}
      \left(-\nabla_1^2-d+(1-\tfrac1\xi)\tfrac{2}{d-1}(W_1\cdot \nabla_1)(K_1\cdot \nabla_1)\right)\Pi(X_1,X_2;W_1,W_2)
    =(W_1\cdot W_2)\delta(X_1,X_2)\,,
    \end{aligned}
\end{equation}
where 
\be\label{eq:Pig01Def}
\Pi\left(X_1, X_2, W_1, W_2\right)=(W_1\cdot W_2) g_0(u)+\left(W_1 \cdot X_2\right)\left(W_2 \cdot X_1\right) g_1(u)\,,
\ee
and $g_{0,1}$ are the two scalar functions to be determined. In ambient space, we have
\begin{equation}
\Pi_{\mu \nu}(x, y)=-g_0(u)\nabla_{\mu}\nabla_{\nu} u+g_1(u)\nabla_{\mu}u\nabla_{\nu}u  \,.
\label{eq:propintr}
\end{equation}
The propagator \eqref{eq:propintr} can also be expressed in terms of the bi-tensors $g_{\mu \nu'}(x,y)$ and $n_\mu(x,y)$ introduced in \cite{Allen1986}, see appendix \ref{app:mapAJ} for the explicit map.

We determine $\Pi$ using the spectral representation, see appendix \ref{app:sprep} for an overview and \cite{Costa:2014kfa} for further details. 
The first point to note is that the transverse part of the gauge field does not depend on $\xi$ and is given by the massless limit of the
first row of the spin 1 bulk propagator in \eqref{eq:Pi1Proca}, with $\Delta = d-1$. We then have 
\begin{equation}\label{eq:EmbPropAns}
\begin{aligned}
    \Pi(X_1,X_2;W_1,W_2)&=\int\! d\nu\, \gamma_1(\nu)\Omega_\nu^{(1)}(X_1,X_2;W_1,W_2)\\
    &\qquad + (W_1\cdot \nabla_1)(W_2\cdot\nabla_2)\int\! d\nu\,\gamma_0(\nu) \Omega_\nu^{(0)}(X_1,X_2)\,,
    \end{aligned}
\end{equation}
where the functions $\Omega_\nu^{(\ell)}$ are defined in \eqref{eq:harmonic}, and
\be 
\gamma_1(\nu) = \frac1{\nu^2+\big(\frac d2-1\big)^2}\,.
\ee
Plugging \eqref{eq:EmbPropAns} into \eqref{eq:spin1P} we then obtain \cite{Ankur:2023lum}
\be
\gamma_0(\nu) = \frac{\xi}{\Big(\nu^2+\frac{d^2}{4}\Big)^2}\,.
\ee
An explicit expression of the propagator is obtained by evaluating the residues of the spectral integrals. As reviewed in appendix \ref{app:sprep}, 
we can get both $D$ and $N$ bulk gauge propagators by an appropriate choice of contour for $\nu$. The $D$ propagator is found by taking $\nu\in (-\infty,\infty)$ and closing the contour at infinity in such a way that the contributions at infinity vanish. This selects the appropriate poles for $\gamma_0$ and $\gamma_1$; the choice of contour that leads to the opposite choice of poles determines the $N$ bulk propagator. 
The explicit form of the $N$ and $D$ bulk propagators in a general $\xi$-gauge is rather involved and is reported in appendix \ref{app:anygauge}. 
Interestingly enough, the $D$ gauge propagator remarkably simplifies for $\xi=d/(d-2)$. In this case, we have
\begin{empheq}[box=\mybluebox]{align}
\begin{split} g_0^{(D)}(u) & =\frac{ \Gamma \left(\frac{d+1}{2}\right)  }{2 \pi ^{\frac{d+1}{2} } (u (u+2))^{\frac{d-1}{2}}(d-2)}\,,\\ 
     g_1^{(D)}(u)& =\frac{u+1} {u (u+2)} g_0^{(D)}(u)\,,\end{split} 
     \qquad \xi=\frac{d}{d-2}\,,
          \label{subYennie}
\end{empheq}
where $g_0^{(D)}$ and $g_1^{(D)}$ are the scalar functions entering \eqref{eq:Pig01Def} and \eqref{eq:propintr} for the $D$ bc. 
A similar simplification occurs for the $N$ bc for the same value of $\xi$,
\begin{empheq}{align}
\begin{split}
        g_0^{(N)}(u)&=\frac{u (u+2) \, _2F_1\left(1,\frac{d}{2};\frac{d+3}{2};-u (u+2)\right) +(d+1)}{4\pi^{d/2} (d+1) (u+1) \Gamma
   \left(2-\frac{d}{2}\right)}+g_0^{(D)}(u)\,,\\
   g_1^{(N)}(u)&=\frac{u+1}{u(u+2)}\left(g_0^{(N)}(u)-\frac{1}{4\pi^{d/2} (u+1) \Gamma
   \left(2-\frac{d}{2}\right)}\right)\,,
    \end{split}\qquad \xi=\frac{d}{d-2}\,.
          \label{subYennieN}
\end{empheq}
For reference, we report both $D$ and $N$ propagators in this gauge in $d=3$, namely
\begin{align}
\begin{split} g_0^{(D,N)}(u)& =\frac{1}{4\pi^2}\left(\frac{1}{u}\mp \frac{1}{u+2}\right)\,,\\
     g_1^{(D,N)}(u)& =\frac{1}{8\pi^2}\left(\frac{1}{u^2}\mp \frac{1}{(u+2)^2}\right)\,,\end{split} 
     \qquad d=3~,~~\xi=3\,,
\end{align}
where the sign $-$ refers to $D$ and $+$ to $N$.

There are two reasons why this gauge choice is special. First, the gauge propagator with $D$ bc enjoys the peculiar
transversality condition
\be\label{eq:TransYennie}
X^B \Pi_{AB}(X,Y) = Y^A \Pi_{AB}(X,Y) = 0\,,  \qquad \xi=\frac{d}{d-2}\,.
\ee
In ambient space, in the basis \eqref{eq:bitensor} of \cite{Allen1986}, the propagator is proportional to $g_{\mu \nu'}+n_\mu n_{\nu'}$ and the condition \eqref{eq:TransYennie} turns into
the transversality\footnote{Recall that while $n^\mu g_{\mu\nu} = n_\nu$, we have $n^\mu g_{\mu\nu'} = - n_{\nu'},n^{\nu'} g_{\mu\nu'} = - n_{\mu}$. Here $g_{\mu\nu}$ is the usual metric tensor, while $g_{\mu \nu'}$ is a bi-tensor. See appendix \ref{app:mapAJ} and \cite{Allen1986} for details.}
\be
n^\mu \Pi_{\mu \nu'} = n^{\nu '} \Pi_{\mu\nu'} = 0\,.
\ee
Second, in flat $d+1$-dimensional space the gauge $\xi=d/(d-2)$ is known as the Fried-Yennie gauge \cite{Fried:1958zz} and is known to lead to a remarkable reduction of IR divergences
in QED, to all orders in perturbation theory \cite{Yennie:1961ad}. Given that AdS can be seen as an IR regulator of flat space, it is perhaps not so surprising that such a gauge leads to remarkable simplifications. We will refer, in what follows, to this gauge as the Fried-Yennie (FY) gauge.

\subsection{Bulk-to-boundary gauge propagator}

The bulk-to-boundary gauge propagator $K_{AB}(X,P)$ can be obtained from the bulk-to-bulk propagator by sending one of the two bulk points to the boundary. 
Note that, for the $D$ bc we also need to divide by a factor $g^2$ to recover the current at the boundary, as expressed in eq.\eqref{eq:BoundLimDN},
\be
\langle J_A(P)\dots\rangle=\lim_{z\rightarrow 0}\frac{1}{g^2} z^{2-d}\langle A_A(X)\dots\rangle\,.
\label{eq:boundlimD}
\ee
Here and in the rest of the section we suppress color indices. In embedding space, we get
\begin{equation}
\begin{aligned}
    K_{AB}^{(D)}(X,P)& =\frac{  \Gamma(d)}{2 (d-2) \pi^{\frac{d}{2}}\Gamma\left(\frac{d}{2}\right)}\frac{(-2P\cdot X)\eta_{AB}+2 P_A X_B}{(-2P\cdot X)^{d}}
    \,.
  \end{aligned}
  \label{eq:boundprop}
\end{equation}
 Note that this propagator does not depend on $\xi$. The bulk-to-boundary $D$ propagator \eqref{eq:boundprop} can also be fixed using exclusively $d$-dimensional conformal invariance at the boundary.

For the $N$ bc, the bulk-to-boundary propagator reads instead
\begin{align}\label{eq:boundpropN}
\begin{split}
     & K_{AB}^{(N)}(X,P)=-g^2\frac{2\Gamma(1+\frac d2)\sin(\frac{\pi d}2)}{\pi^{\frac {d+2}2}(d-2)^2d}\left(\frac{(d-1)(-2P\cdot X) \eta_{AB}+2P_AX_B}{(-2P\cdot X)^2}\right)\\
    &\quad -g^2 \zeta \frac{2\Gamma(1+\frac d2)\sin(\frac {\pi d}2)}{\pi^{\frac{d+2}{2}}d^2}\left[C(d)+\log \left(-{2P\cdot
   X}\right)\right]\left(\frac{(-2P\cdot X)\eta_{AB}+2P_A X_B}{(-2P\cdot X)^2}\right) \\
   &\quad+ g^2 \zeta\frac{4\Gamma(1+\frac d2)\sin(\frac {\pi d}2)}{\pi^{\frac{d+2}2}d^2}\frac{P_A X_B}{(-2X\cdot P)^2}\,,
\end{split}
\end{align}
where we have introduced the shifted gauge-fixing parameter
\be
\zeta=\xi-\frac d{d-2}\,,
\ee
which vanishes in FY gauge, and 
\be 
C(d)=\pi  \cot \left(\tfrac{\pi  d}{2}\right)+2 \psi(d)-\psi \left(\tfrac{d+1}{2}\right)+\gamma_E-\log 4\,,
\ee
 $\psi$ being the digamma function and $\gamma_E$ being the Euler-Mascheroni constant. Note the appearance of log terms in a generic gauge and how 
 also the bulk-to-boundary $N$ propagator simplifies considerably in the FY gauge.
 The presence of $\zeta$-dependent terms is due to the fact that the corresponding boundary operator, the gauge connection $a$, is not gauge invariant. On the other hand, the bulk-to-boundary propagator for the field strength,
 \begin{align}
     \begin{split}
         \langle F_{AB}(X)f_{CD}(P)\rangle = \frac{4g^2 \Gamma(\frac d2)\sin(\frac{\pi d}2)(\calP_{AC}(X,P)\calP_{BD}(X,P)-\calP_{AD}(X,P)\calP_{BC}(X,P))}{\pi^{\frac{d+2}2}(d-2)(-2P\cdot X)^2}\,,
     \end{split}
 \end{align}
 where we have introduced the projector
 \begin{equation}\label{eq:projc}
     \calP_{AB}(X,P)=\eta_{AB}+\frac{2P_A X_B}{(-2P\cdot X)}\,,
 \end{equation}
 is $\zeta$-independent and has the appropriate structure for the bulk-to-boundary correlator of an antisymmetric rank-2 tensor.
Contrary to the $D$ case, \eqref{eq:boundpropN} has a factor $g^2$, because the boundary limit in \eqref{eq:BoundLimDN} does not require to divide by $g^2$ in this case. 
\subsection{Ghost propagator}

The two possible boundary conditions for the ghost fields $c$ are $c(x,z)\underset{z\to 0}{\sim} z^{\Delta} \,\hat{c}_{\Delta}(x)$, with either $\Delta=0$ or $\Delta=d$.
 They are constrained by the choice of boundary condition on the gauge fields: with $N$ bc, the presence of dynamical gauge fields at the boundary requires the gauge transformation (and equivalently the ghost field) to persist at the boundary, i.e. $\Delta = 0$; with $D$ bc, the gauge transformations should instead decay faster than the gauge fields at the boundary, as the bulk gauge fields are dual to global currents in this case, therefore the correct bc is $\Delta =d$. 
 The ghost propagator with $D$ bc $G_{{\rm GH}}^{ (D)}$ is simply the propagator of a massless scalar field,
 \begin{equation}
G_{{\rm GH}}^{ (D)}(X,Y)= \frac{\Gamma\left(\frac{d+1}{2}\right)}{2d\pi^{\frac{d+1}{2}}u^{d}}\,{ }_2 F_1\left(d, \frac{d+1}{2},d+1,-\frac{2}{u}\right)\,.
\label{eq:GHprop}
 \end{equation}
The $N$ ghost propagator can similarly be derived by applying the prescription \eqref{eq:scalarN},
 \begin{equation}\begin{aligned}
 G_{{\rm GH}}^{ (N)}(X,Y)&= -\frac{_3F_2\left(1,1,\frac{3}{2};2,2-\frac{d}{2};\frac{1}{(1+u)^2}\right)}{4\pi ^{d/2} (d-2) (1+u)^2 \Gamma \left(1-\frac{d}{2}\right)}\\
 &\qquad+\frac{  \psi\left({\frac{1-d}{2}}\right)-2\psi({1-d})-\gamma_E-\log \left(\frac{1+u}{2}\right)-\frac1d}{2 \pi ^{d/2}  \Gamma
   \left(1-\frac{d}{2}\right)}\,,\end{aligned}
\label{eq:GHpropN}
 \end{equation}
where $\psi(z)$ is the digamma function and $\gamma_E$ is the Euler-Mascheroni constant.

\section{Anomalous dimensions of lightest scalar operators}
\label{sec:gammaJJ}

In this section, we compute the anomalous dimension of the lightest scalar singlet boundary operator, in both the $D$ and the $N$ bc's. The operator is $\mathrm{tr}[J_i J^i]$ for $D$ bc, and  $\mathrm{tr}[f_{ij} f^{ij}]$ for $N$ bc, where $f_{ij} = \partial_i a_j -\partial_j a_i - i [a_i,a_j]$. As already mentioned in the introduction, for both cases in the limit $g^2\to 0$ this operator has dimension 4 and it coincides (up to a normalization factor) with the displacement operator of the theory at the free UV fixed point. The latter statement can be proved either by using the expression for the bulk stress tensor and taking the boundary OPE (bOPE) limit, or by noticing that it is the only singlet dimension 4 operator in the boundary spectrum (see table \ref{tab:doubletraceprimaries}), and therefore the only candidate to be the displacement operator, which must exist in the spectrum when the bulk is a CFT. 

To do the computation we will first exploit a multiplet-recombination argument, that fixes the leading-order anomalous dimension of the displacement operator for any perturbation of a CFT in AdS background. We present the argument in this general setting in subsection \ref{subsec:displop}.  We then discuss what this result teaches us regarding the disappearance/persistence of the $D$/$N$ bc as we increase the AdS radius. For the case of $D$ bc, we then check the result with an explicit diagrammatic calculation in subsection \ref{subsec:JJ2pt}. 

\subsection{Anomalous dimension of the displacement operator}
\label{subsec:displop}

A CFT in AdS$_{d+1}$ is equivalent up to a Weyl rescaling to a BCFT. A general BCFT result then implies that any CFT in AdS$_{d+1}$ must have a boundary operator $\mathcal{D}$ of dimension $\Delta_\mathcal{D}=d+1$, which appears in the bOPE of the bulk stress tensor \cite{Liendo:2012hy}. This operator is the so-called displacement operator. The two-point function between the traceless bulk stress tensor and the boundary displacement operator is fixed by the isometries. In embedding space, it reads
\begin{equation}\label{eq:DisT2pt}
\text{CFT:}\quad\langle T_{AB}(X) \mathcal{D}(P) \rangle = \frac{C_{T\mathcal{D}}}{(-2 P\cdot X)^{d+1}}\left(\frac{G_{AC}(X)G_{BD}(X) P^C P^D}{(-2 P\cdot X)^2}- \frac{G_{AB}(X)}{4(d+1)}\right)\,,
\end{equation}
where $G_{AB}(X)= \eta_{AB}+X_A X_B$ is the projector to the tangent space at $X$. The two-point function \eqref{eq:DisT2pt} is fixed up to normalization by the two requirements 
\begin{align}
&\nabla_X^A \langle T_{AB}(X) \mathcal{D}(P) \rangle = 0\,,\\
& G^{AB}(X) \langle T_{AB}(X) \mathcal{D}(P) \rangle = 0\,,
\end{align}
where the first is the conservation, and the second is the traceless condition appropriate to a CFT. Assuming the bulk stress tensor is normalized via the Ward identities for the isometries, the coefficient $C_{T\mathcal{D}}$ 
depends on the normalization of the operator $\mathcal{D}$. For the sake of our argument, we can leave the precise choice of normalization unspecified.

Next, we turn on a deformation in the bulk, i.e.
\begin{equation}
S_{\text{bulk}} = S_{\text{CFT}} + \lambda \int dx\, O(x)\,,
\end{equation}
where $O$ is an operator of scaling dimension $\Delta_O$ of the bulk CFT. As a consequence of the deformation the stress tensor acquires a non-zero trace. In embedding space we have
\begin{equation}\label{eq:Trrel}
\Delta_O \neq d+1:\quad G^{AB}(X) T_{AB}(X) = (\Delta_O -d-1)\lambda\, O(X) + \alpha(\lambda) \mathbf{1}\,.
\end{equation}
Besides the operator violation of scale-invariance proportional to the deformation $O$, we also allow a $c$-number contribution proportional to the identity operator, which is generally present due to the curvature of the background, with a coefficient $\alpha(\lambda)$ that depends on the deformation $\lambda$.\footnote{In the special case of $d+1=4$, this coefficient is a linear combination of beta functions for curvature terms \cite{Brown:1980qq,Hathrell:1981zb,Hathrell:1981gz,Freeman:1983cx}, with couplings denoted by $a$, $b$ and $c$. There is no need to specify their form since, as we will see, they will not play any role in our analysis.}
In the special case $\Delta_O=d+1$ the coupling $\lambda$ is classically marginal and we have instead (assuming for simplicity that there is a single marginal operator in the CFT)
\begin{equation}\label{eq:Trmarg}
\Delta_O = d+1: \quad G^{AB}(X)  T_{AB}(X) = \beta_\lambda(\lambda)\, O(X)+ \alpha(\lambda) \mathbf{1}\,.
\end{equation}
Here $\beta_\lambda$ is the beta function, which for small $\lambda$ behaves as 
\begin{equation}
\beta_\lambda(\lambda) = \beta_0 \,\lambda^{n} + \mathcal{O}(\lambda^{n+1})\,,
\end{equation}
for some integer $n>1$ and some real coefficient $\beta_0$.\footnote{As pointed out in \cite{Jack:1983sk,Osborn:1991gm}, in $d+1=4$ and in presence of continuous global symmetries, the beta functions $\beta_i$ are subject to a possible ambiguity and are replaced by well-defined functions $B_i$. This issue will not appear in the YM application as there are no
continuous global symmetries.}

As a consequence of the trace being non-zero, there is an additional structure in the two-point function, and the dimension $\Delta_{\mathcal{D}}(\lambda)$ of the operator $
\mathcal{D}$ will depend on $\lambda$ and no longer be protected, so the two-point function is  
\begin{align}\label{eq:2ptTDdef}
\begin{split}
\text{$\lambda>0$:}\quad\langle T_{AB}(X) \mathcal{D}(P) \rangle  = \frac{C_{T\mathcal{D}}(\lambda)}{(-2 P\cdot X)^{\Delta_{\mathcal{D}}(\lambda)}}&\left(\frac{G_{AC}(X)G_{BD}(X) P^C P^D}{(-2 P\cdot X)^2}- \frac{G_{AB}(X)}{4(d+1)}\right. \\
&\qquad \left. - \frac{(\Delta_{\mathcal{D}}(\lambda)-d-1)d}{4(d+1)\Delta_{\mathcal{D}}(\lambda)}  G_{AB}(X) \right)\,.
\end{split}
\end{align}
The coefficient of the additional structure in the second line is fixed in terms of $\Delta_{\mathcal{D}}$ and $C_{T\mathcal{D}}$ once we impose the conservation of the stress tensor. Note that besides $\Delta_{\mathcal{D}}$ also the normalization $C_{T\mathcal{D}}$ acquires a dependence on $\lambda$ as we have indicated. The two-point correlator between the deformation $O$ and the displacement is fixed by bulk isometries to have the form
\begin{equation}\label{eq:OD2pt}
\langle O(X) \mathcal{D}(P) \rangle = \frac{C_{O\mathcal{D}}(\lambda)}{(-2 P\cdot X)^{\Delta_{\mathcal{D}}(\lambda)}}\,.
\end{equation}
Taking the trace of equation \eqref{eq:2ptTDdef}, using the operator equation \eqref{eq:Trrel} and substituting \eqref{eq:OD2pt} we obtain the relation
\begin{equation}\label{eq:DisNoMargEx}
 \Delta_O\neq d+1: \quad \Delta_{\mathcal{D}}(\lambda)-d-1 = - \frac{4(\Delta_O-d-1)\lambda}{d} \frac{C_{O\mathcal{D}}(\lambda)}{C_{T\mathcal{D}}(\lambda)}\Delta_{\mathcal{D}}(\lambda)\,.
\end{equation}
Note that the $c$-number contribution given by $\alpha(\lambda)$ in \eqref{eq:Trrel} drops from the two-point function because it gives rise to a 
one-point function for the boundary operator $\mathcal{D}$, which vanishes.
Expanding this expression at small $\lambda$ and denoting $\Delta_{\mathcal{D}}(\lambda)-d-1 = \gamma_{\mathcal{D}}(\lambda)$ we obtain that the leading order anomalous dimension of the displacement operator is
\begin{equation}\label{eq:anodimDgen}
\Delta_O\neq d+1: \quad \gamma_{\mathcal{D}}(\lambda) = - \frac{4 (d+1)(\Delta_O-d-1)}{d} \frac{C_{O\mathcal{D}}}{C_{T\mathcal{D}}}\lambda + \mathcal{O}(\lambda^2)\,.
\end{equation}
When the $\lambda$ dependence is not explicitly indicated in the normalization coefficients $C_{T\mathcal{D}}$ and $C_{O\mathcal{D}}$, we mean their values at $\lambda= 0$, i.e. in the CFT. Note that the normalization choice for $\mathcal{D}$ does not matter in this formula because it cancels in the ratio between normalization coefficients. The relative normalization between $T$ and $\lambda \,O$ on the other hand is fixed by the operator equation \eqref{eq:Trrel}. In the special case of a classically marginal deformation with $\Delta_O = d+1$, following the same steps and using the operator equation \eqref{eq:Trmarg} instead, we obtain the relation
\begin{empheq}[box=\mybluebox]{align}\label{eq:CoCTD3}
 \Delta_O= d+1: 
 \quad \Delta_{\mathcal{D}}(\lambda)-d-1 = - \frac{4\beta_\lambda(\lambda)}{d} \frac{C_{O\mathcal{D}}(\lambda)}{C_{T\mathcal{D}}(\lambda)}\Delta_{\mathcal{D}}(\lambda)\,,
\end{empheq}
which is valid to all orders in perturbation theory. The same remark applies as well to \eqref{eq:DisNoMargEx}.
Expanding \eqref{eq:CoCTD3} at small $\lambda$ gives the following result for the leading order anomalous dimension
\begin{empheq}[box=\mybluebox]{align} \label{eq:anodimDmarg}
\Delta_O =  d+1: \quad \gamma_{\mathcal{D}}(\lambda) = - \frac{4 (d+1)}{d} \frac{C_{O\mathcal{D}}}{C_{T\mathcal{D}}} \,\beta_0 \,\lambda^n + \mathcal{O}(\lambda^{n+1})\,.
\end{empheq}
Therefore, in the presence of a classically marginal running coupling in the bulk, the leading anomalous dimension of the displacement is fixed by the leading coefficient in the beta function of the bulk coupling.

\subsubsection{Application to YM}
YM theory in AdS$_4$ does not fall straightforwardly in the setup described above of a CFT with a small deformation. 
For definiteness, we discuss $SU(n_c)$ YM, the generalization to other gauge groups is straightforward.
At the level of local operators, the UV CFT is the abelian theory of $n_c^2-1$ free gluons (see e.g. \cite{Antinucci:2022eat} for a discussion of the global structure of the theory in this limit). The deforming operator is the Lagrangian itself, with a large coefficient $\frac{1}{g^2}$. Nevertheless, this can be treated perturbatively because of the factors of $g^2$ in each gluon propagator.

The stress tensor of YM theory is
\begin{equation}
T_{\mu\nu} = \frac{2}{g^2}\mathrm{tr}\left[F_\mu^{\phantom{\mu}\rho}F_{\nu\rho} - \frac{g_{\mu\nu}}{4} F^{\rho\sigma}F_{\rho\sigma}\right]\,.
\end{equation}
Its trace is given by 
\begin{equation}
T^\mu_{\phantom{\mu}\mu} = \beta_\frac{1}{g^2} \mathrm{tr}\left[\frac{1}{2}F^{\rho\sigma}F_{\rho\sigma}\right]+\alpha(g^2)\mathbf{1} = -\frac{1}{g^4} \beta_{g^2}\mathrm{tr} \left[\frac{1}{2}F^{\rho\sigma}F_{\rho\sigma}\right]+\alpha(g^2)\mathbf{1}\,.
\end{equation}
Like in the previous section, we allowed a $c$-number contribution with a $g^2$-dependent coefficient, whose form has been first determined in \cite{Freeman:1983cx}. This contribution drops from the anomalous dimension. The one-loop beta function is
\begin{equation}
\beta_{g^2}(g^2) = - \frac{22 n_c}{3}\frac{g^4}{(4\pi)^2} + \mathcal{O}(g^6)\,.
\end{equation}
In the notation of the previous section, calling 
\begin{equation}
O= -\frac{1}{g^2}\mathrm{tr} \left[\frac{1}{2}F^{\rho\sigma}F_{\rho\sigma}\right]\,,
\end{equation}
eq.~\eqref{eq:anodimDmarg} gives
\begin{equation}\label{eq:anodimDYM}
\gamma_{\mathcal{D}}(g^2)= \frac{16}{3} \frac{C_{O\mathcal{D}}}{C_{T
\mathcal{D}}} \frac{22 n_c}{3}\frac{g^2}{(4\pi)^2}+\mathcal{O}(g^4)\,.
\end{equation}

We now specify the boundary conditions and compute the coefficients that enter the anomalous dimension, which are given by the diagrams in figure \ref{fig:CoeffD}. For the $D$ bc, we have $\mathcal{D} =g^2\,\mathrm{tr}[J_i J^i]$ and using the propagator \eqref{eq:boundprop} we get
\begin{align}\label{eq:CTDCOD}
D:\quad\begin{split}
C_{T\mathcal{D}} & = (n_c^2 -1)\frac{32}{\pi^4 } \,,\\
C_{O\mathcal{D}} & = - (n_c^2 -1)\frac{6}{\pi^4 }\,.
\end{split}
\end{align}
Substituting in \eqref{eq:anodimDYM} we obtain
\begin{empheq}[box=\mybluebox]{align}
D:\quad \gamma_{\mathrm{tr}[JJ]}(g^2) = -\frac{11 n_c}{24\pi^2} g^2 +\mathcal{O}(g^4)\,.
\label{eq:gammaJJdisp}
\end{empheq}
For the $N$ bc, on the other hand, $\mathcal{D}=-\frac1{2g^2}\mathrm{tr}[f_{ij}f^{ij}]$. Using the propagator \eqref{eq:boundpropN} we get
\begin{align}
N:\quad\begin{split}
C_{T\mathcal{D}} & =  (n_c^2 -1)\frac{32}{\pi^4} \,,\\
C_{O\mathcal{D}} & =  (n_c^2 -1)\frac{6}{\pi^4}\,.
\end{split}
\end{align}
Substituting in \eqref{eq:anodimDYM} we obtain
\begin{empheq}
[box=\mybluebox]{align}
N:\quad \gamma_{\mathrm{tr}[ff]}(g^2) =  \frac{11 n_c}{24\pi^2} g^2 +\mathcal{O}(g^4)\,.
\label{eq:gammaJJdispN}
\end{empheq}
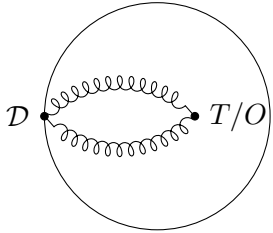
\begin{figure}
\centering
\begin{tikzpicture}
  \begin{feynman}[inline=(a.base)]
  \draw (0,0) circle (1.5);
    \vertex [dot](a) at (-1.5,0) {\small{}}; 
    \node [left] at (-1.5,0) {$\mathcal{ D}\,$};
    \vertex [dot] (b) at (1/2,0){\small{}}; 
    \node [right] at (1/2,0){$\,T/O$};

    \diagram* {
      (b)-- [gluon,quarter left] (a),
      (a) -- [gluon, quarter left] (b)
    };
  \end{feynman}
\end{tikzpicture} 
\caption{Diagrams that compute the coefficients $C_{O\mathcal{D}}$ and $C_{T\mathcal{D}}$ in the free UV limit $g\to 0$.} \label{fig:CoeffD}
\end{figure}
Interestingly enough, the leading corrections \eqref{eq:gammaJJdisp} and \eqref{eq:gammaJJdispN} are equal and opposite. We do not know if this is a mere coincidence of the leading 
contribution or if there is some mechanism explaining this relation. It would be interesting to better understand this point.

As we mentioned in the introduction, if a boundary singlet scalar operator is marginal for some value of $g^2$, or equivalently for some value of $\Lambda_{\text{YM}}$ in units of the AdS radius, then the corresponding boundary condition goes through merger and annihilation \cite{Kaplan:2009kr,Gorbenko:2018ncu} and it stops existing as a unitary boundary condition. This phenomenon, first envisioned in \cite{Hogervorst:2021spa}, was explained in detail in \cite{Lauria:2023uca} and it was applied to two-dimensional theories in \cite{Copetti:2023sya}. The leading order anomalous dimensions obtained above are suggestive that the displacement operator indeed becomes marginal for the $D$ bc and not for the $N$ bc. This matches the expectation that the $D$ bc should not exist for arbitrary large AdS because it would give rise to massless and colored asymptotic states in flat space, while it is possible that the $N$ bc approaches smoothly the flat space limit. Truncating the scaling dimension of $\mathrm{tr}[JJ]$ to the leading order correction \eqref{eq:gammaJJdisp} gives the estimate \eqref{eq:estimate} for the transition in the Marginality scenario.

Let us briefly review, in our setup, the derivation of the merger and annihilation presented in \cite{Lauria:2023uca}. Assume that, as suggested by the calculations above, at a certain value $g^2 = g^2_{\text{crit}}$ the operator $\mathrm{tr}[JJ]$ is marginal. Whenever we have a marginal operator, it is a nontrivial requirement that the  $\beta$ function of the associated boundary coupling has real zeroes, and we need to impose it in order to preserve conformal symmetry on the boundary. This $\beta$ function can be calculated reliably in a double expansion, in the marginal coupling itself, and in the displacement of the bulk coupling $1/g^2$ from the critical value $1/g^2_c$. To avoid confusion, we stress that this is not the same perturbation theory that we used above to obtain the anomalous dimensions of $\mathcal{D}$, the latter being an expansion around the free CFT in the bulk i.e. $g^2\ll1$, while now we are interested in expanding about an interacting bulk theory. In this double expansion, the beta function for the boundary marginal coupling $\eta \, \mathrm{tr}[JJ]$ is given by 
(see appendix \ref{app:boundaryRG} or \cite{Lauria:2023uca}) 
\begin{equation}\label{eq:betamerging2}
\beta_\eta = c_1 \eta^2 + c_2 \left(\frac{1}{g^2}-\frac{1}{g^2_\text{crit}}\right)+\text{subleading}\,,
\end{equation}
where (plugging $d=3$ in the formulas of appendix \ref{app:boundaryRG})
\be
c_1  = 2\pi \frac{C_3}{C_2}\Big|_{g^2=g^2_{\text{crit}}}  \,, \qquad\qquad
c_2  =-\frac{B}{C_2}\Big|_{g^2=g^2_{\text{crit}}} \,,
\ee
with $C_2$ and $C_3$ the two- and three-point function coefficients of $\mathrm{tr}[JJ]$, and $B$ the coefficient of the two-point function between the bulk Lagrangian {$\mathrm{tr}[\frac{1}{2}F_{\mu\nu}F^{\mu\nu}]$} and $\mathrm{tr}[JJ]$:
\begin{equation}
\left\langle \mathrm{tr}[\tfrac{1}{2}F_{\mu\nu}F^{\mu\nu}](X) 
 \mathrm{tr}[JJ](P)\right\rangle = \frac{B(g^2)}{(-2 P\cdot X)^d}\,.
\end{equation}
At leading order for $g^2\ll 1$, $C_2>0$ is given by \eqref{eq:CJJ} below, $C_3 >0$,
and $B = -g^2 C_{O\mathcal{D}}>0$, with
$C_{O\mathcal{D}}$ given in \eqref{eq:CTDCOD}. The value of these coefficients at $g^2_{\text{crit}}$ is beyond the reach of perturbation theory.
Nevertheless, the assumption that the $D$ bc is a viable unitary boundary condition that preserves AdS isometries in the range of coupling $0\leq g^2 \leq g^2_{\text{crit}}$
guarantees that $\beta_\eta$ must have real zeros for $g^2\lesssim g^2_\text{crit}$. This implies that $c_1$ and $c_2$ must have opposite signs.
As a result $\beta_\eta$ has two real zeros, at
$\eta_\pm =\pm \sqrt{c_2/c_1 (g_\text{crit}^{-2}-g^{-2})}$. So, at least close to $g^2_{\text{crit}}$, another boundary condition $D^*$ must exist, which 
gives rise to another boundary CFT, with the same global symmetry of the $D$ CFT, namely $SU(n_c)$. To the operator $O_+ = \mathrm{tr}[JJ]$ of the $D$ theory is associated another singlet scalar operator $O_-$ of the $D^*$ theory. Their dimensions are determined by the slope of the $\beta$ function at the corresponding zero, giving \cite{Gorbenko:2018ncu} 
\be
\Delta_\pm = d + 2 |c_1| \eta_\pm  \,.
\ee
Deforming the $D^*$ CFT with $O_-$ leads to a (short) RG flow ending in the CFT $D$. When $g^2=g^2_\text{crit}$, the two CFTs merge and annihilate, namely they 
turn to complex CFT for $g^2>g^2_{\text{crit}}$, with purely imaginary anomalous dimensions for $O_\pm$ close to the merging point \cite{Gorbenko:2018ncu}. 

Note that this mechanism of loss of conformality has been advocated in \cite{Kaplan:2009kr} as a possible explanation of how
conformal windows terminate in $4d$ non-abelian gauge theories with matter. In that context, the role of $g^{-2}$ is played by the number of flavours.\footnote{In the Veneziano limit the number of flavors is replaced by a continuous parameter and the merge and annihilation scenario can be analyzed in controlled set-ups, see e.g. \cite{Benini:2019dfy}. In this limit, the role of $\eta$ is played by a double trace deformation.}
Interestingly enough, here we are advocating the possibility that confinement itself in pure Yang-Mills theory can be explained as a mechanism of loss of conformality, but this time the CFT in question is a 3d CFT living at the boundary of AdS space. 

It is also interesting to observe that a similar instability of the $D$ bc exists in three-dimensional gauge theories in AdS$_3$. In that case however the singlet scalar operator bilinear in the currents is actually marginal at zero bulk coupling, causing the Dirichlet boundary condition to be unavailable already in perturbation theory, see e.g. \cite{Faulkner:2012gt, Ankur:2023lum}.

\subsection{Anomalous dimensions from $J^aJ^b$ two-point function }
\label{subsec:JJ2pt}

We now restrict to $D$ bc and compute the anomalous dimensions of the lightest scalar primaries with different representations of $SU(n_c)$. To do this, we perform a direct computation of Witten diagrams contributing to the two-point functions of $J^aJ^b(x)=J^{ia}J_{i}^{\ b}(x)$. Here vector indices are contracted, while color indices are left open. From now on we drop the superscript $D$ in Dirichlet propagators, as $N$ bc will no longer enter our discussion. 

The operator $J^aJ^b$ is the symmetric product of two fields in the adjoint, which decomposes into irreducible representations according to
\begin{equation}\label{eq:rappsymm}
R_A\otimes R_A |_\text{sym}=R_+= \mathbf{1} \oplus R_A \oplus R_1 \oplus R_3\,,
\end{equation}
where $\mathbf{1}$ is the singlet, $R_A$ is the adjoint, and $R_1$, $R_3$ are defined in appendix \ref{app:characters}.\footnote{This decomposition is valid only for $n_c>3$, see appendix \ref{app:characters} for details on the decomposition for $n_c=2,3$.} Note that the singlet corresponds to the  operator tr$[JJ]$, which we studied in the previous subsection. Matching \eqref{eq:gammaJJdisp} with the result obtained with a direct computation will be a non-trivial check of our computations. 

Given a scalar primary operator $O$ of classical dimension $\Delta_O^{(0)}$, its two-point function is given by
  \begin{equation}
  \langle O(x) O(0) \rangle = \frac{c_O(g^2)}{x^{2\Delta_O}} \,,
     \label{eq:2ptO}
 \end{equation}
 where $\Delta_O = \Delta_O^{(0)}+\gamma_O$. In perturbation theory we can  determine the leading-order anomalous dimension $\gamma_O$ by looking at
 the logarithmic part of the next-to-leading correction to the two-point function:
 \begin{equation}
     \gamma_O(g^2) =-\frac{\langle \left.O(x_1)O(x_2)\rangle^\text{1}\right|_{\text{log}\, x^2}}{\langle O(x_1)O(x_2)\rangle^0}+\mathcal{O}(g^4)\,.
     \label{eq:gammaprim}
 \end{equation}
 Here and in the following we use the superscript 0 and 1 to denote the leading order and the next-to-leading order respectively.

We  first compute the leading order and the logarithmic terms at the next-to-leading order of the two-point function  $\langle J^{a_1}J^{b_1}(x_1) J^{a_2}J^{b_2}(x_2)\rangle$ and then project onto irreducible representations. 
At the leading order, the two-point function in $d=3$ reads 
 \begin{equation}
 \langle J^{a_1}J^{b_1}(x_1) J^{a_2}J^{b_2}(x_2)\rangle^0 = \frac{C_{JJ}^0}{g^4}   (\delta^{a_1 a_2}\delta^{b_1 b_2} +\delta^{a_1 b_2}\delta^{b_1 a_2})   \frac1{x_{12}^{2\Delta^0_{JJ}}}\,,
 \label{eq:JJ}
 \end{equation}
 with 
 \begin{equation}
 C_{JJ}^0=3(C^0_J)^2=\frac{12}{\pi^4},\qquad \Delta_{JJ}^0=2\Delta_{J}=4\,,
 \label{eq:CJJ}
 \end{equation}
 where  $C^0_J$ is defined in \eqref{eq:CJ0}.
Let us now compute the next-to-leading order.
 We consider the bulk two-point function of the composite operator $A^{\mu a}A_\mu^{b}$,
 \begin{equation}
g^4 Q(x_1,x_2)=\langle A^{\mu_1 a_1} A_{\mu_1}^{b_1}(x_1) A^{\mu_2 a_2} A_{\mu_2}^{b_2}(x_2) \rangle\,.
 \end{equation}
We then uplift it to embedding coordinates and take the external points to the boundary to compute $Q^{\partial} (P_1,P_2)$.
While $Q$ is not gauge invariant, the $\xi$-dependence cancels when we take the boundary limit, making $Q^\partial$ gauge invariant instead.

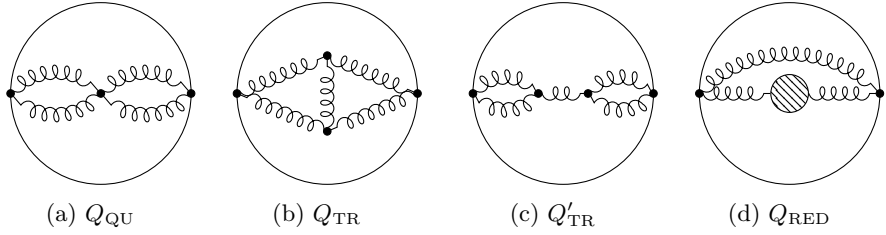
\begin{figure}
  \centering 

            \begin{subfigure}[c]{0.14\textwidth}
                    \begin{tikzpicture}
  \begin{feynman}[inline=(a.base)]
  \draw (0,0) circle (1.2);
  \vertex [dot] (a) at (0,0){\small{ }};
    \vertex [dot](i1) at (-1.2,0) {\small{ }} ;
    \vertex [dot] (b) at (1.2,0){\small{ }};

    \diagram* {
      (i1)-- [gluon,quarter left] (a),
      (a) -- [gluon, quarter left] (i1),(b)-- [gluon,quarter left] (a),
      (a) -- [gluon, quarter left] (b)
    };
  \end{feynman}
\end{tikzpicture} 
        \caption{$Q_{\rm QU}$}
        \end{subfigure}\qquad \begin{subfigure}[c]{0.14\textwidth}
                    \begin{tikzpicture}
  \begin{feynman}[inline=(a.base)]
  \draw (0,0) circle(1.2);
    \vertex [dot](i1) at (-1.2,0){\small{ }} ;
    \vertex[dot](i2) at (0,0.5){\small{ }} ;
    \vertex [dot] (b) at (1.2,0){};
    \vertex [dot] (o2) at (0,-0.5){} ;

    \diagram* {
      (i1)-- [gluon] (i2)-- [gluon] (b)-- [gluon] (o2) -- [gluon] (i1),(o2)--[gluon](i2)
    };
  \end{feynman}
\end{tikzpicture}\caption{$Q_{\rm TR}$} \end{subfigure} \qquad
        \begin{subfigure}[c]{0.14\textwidth}\centering
                    \begin{tikzpicture}
  \begin{feynman}[inline=(a.base)]
  \draw (0,0) circle (1.2);
  \vertex [dot] (a) at (-0.33,0){};
    \vertex [dot](i1) at (-1.2,0){\small{ }} ;
    \vertex [dot] (b) at (0.33,0){};
    \vertex [dot] (o2) at (1.2,0){\small{  }};

    \diagram* {
      (i1)-- [gluon,quarter left] (a),
      (a)-- [gluon,quarter left] (i1),
      (a)--[gluon] (b),
      (o2)-- [gluon,quarter left] (b),
      (b)-- [gluon,quarter left] (o2)
    };
  \end{feynman}
\end{tikzpicture}

        \caption{$Q_{\rm TR}'$}
        \end{subfigure}\qquad \begin{subfigure}[c]{0.14\textwidth}\begin{tikzpicture}
  \begin{feynman}[inline=(a.base)]
  \draw (0,0) circle(1.2);
  \vertex (a) at (0,0) [blob]{};
    \vertex [dot](i1) at(-1.2,0){\small{ }} ;
    \vertex [dot] (b) at (1.2,0){};

    \diagram* {
      (i1)-- [gluon,quarter left](b),(i1)--[gluon](a)--[gluon](b)
    };
  \end{feynman}
\end{tikzpicture} 
        
        \caption{$Q_{\rm RED}$}
        \end{subfigure}   \caption{Next-to-leading corrections to the two-point function of $J^aJ^b$: the quartic (a), the  triple (b), and the reducible (d) diagram. The diagram (c) evaluates to zero since it has a vanishing color structure and the diagram (d), being free of logarithmic terms,  does not affect the anomalous dimension.}
        \label{fig3}
    \end{figure}
  At next-to-leading order, $Q(x_1,x_2)$ is given by three contributions, corresponding to the diagrams depicted in figure \ref{fig3},
 \begin{equation}
 Q^{(1)}(x_1,x_2)=Q_{\text{QU}}(x_1,x_2)+ Q_{\text{TR}}(x_1,x_2)+ Q_{\text{RED}}(x_1,x_2)\,.
 \label{eq:qtwo}
 \end{equation}
We call these diagrams the quartic, the triple, and the reducible diagram respectively. By reducible here we refer to the fact that the last diagram is simply the contraction between the vector propagator at tree level and its one-loop correction. This contribution can be disregarded, as currents are protected by gauge symmetry and do not get an anomalous dimension. Let us instead focus on the first two terms of the sum in \eqref{eq:qtwo}.

\subsubsection{Quartic diagram}

The quartic diagram is easier to deal with, as it involves integration over only one point,
\begin{equation}
\begin{aligned}
    Q_{\text{QU}}(x_1,x_2)=-g^2 \sigma \intads dx\ \big(&\Pi^{\mu_1 \nu}\Pi_{\mu_1 \nu}(x_1,x)\Pi^{\lambda \mu_2}\Pi_{\lambda \mu_2}(x,x_2)\\ &-\Pi^{\mu_1 \nu}\Pi_{\mu_1 \lambda}(x_1,x)\Pi^{\lambda \mu_2}\Pi_{\nu \mu_2}(x,x_2)\big)\,,
    \end{aligned}
         \label{eq:quemb}
\end{equation}
where we have factored out the color structure 
\be
 \sigma^{a_1 a_2 b_1 b_2} \equiv f^{a_1a_2 c}f^{b_1 b_2}{}_{c} + f^{a_1b_2 c}f^{b_1 a_2}{}_{c}\,. \label{T main text}
\ee
Uplifting to embedding space and taking the boundary limit as in \eqref{eq:boundlimD},
we have
\begin{equation}
\begin{aligned}
Q^{\partial}_{\mathrm{QU}}(P_1,P_2) =-\frac{1}{g^2} \sigma\int dX&(K^{A_1 B}K_{A_1 B}(P_1,X)K^{C A_2}K_{C A_2}(X,P_2)\\ &-K^{A_1 B}K_{A_1 C}(P_1,X)K_{B A_2}K^{C A_2}(X,P_2))\,,
\end{aligned}
\end{equation}
where for simplicity from now on we suppress the color indices in $\sigma$.
We then plug in the expression for the bulk-to-boundary propagator in \eqref{eq:boundprop} to get
\begin{equation} \label{eq:qufin}
\begin{aligned}
Q^{\partial}_{\mathrm{QU}}(P_1,P_2) &=\frac{1}{g^2}  \sigma \frac{ 2^{4d}\Gamma(\frac{d+1}{2})^4\pi^{2d+1} }{16(d-2)^4}    \\ 
&\int dX  \frac{(P_1\cdot P_2)^2 +   2 (P_1\cdot P_2)( P_1\cdot X )(P_2\cdot X )- d(
     d-1) (P_1\cdot X)^2 (P_2\cdot X)^2}{\left(-2P_1 \cdot X\right)^{2d}\left(-2P_2 \cdot X\right)^{2d}} \,.
     \end{aligned}
  \end{equation}
This is a linear combination of scalar integrals of the form
\be
\mathcal{I}_\Delta=\int d X \frac{1}{\left(-2P_1 \cdot X\right)^\Delta} \frac{1}{\left(-2P_2 \cdot X\right)^\Delta}\,,
\label{eq:massshift}
\ee  
which evaluates to
\be
\mathcal{I}_\Delta=
\frac{ \pi^{\frac{
  d}{2}}  \Gamma\left(\Delta-\frac{d}{
     2}\right) }{\Gamma(\Delta)(-2 P_1\cdot P_2)^\Delta}\left(\log\Big(\frac{-2 P_1\cdot P_2}{\delta^2}\Big) - 
   \psi( \Delta)+ 
    \psi\left( 1 - \frac{d}{2} + \Delta\right)\right)\,,
    \label{eq:massshiftresult}
\ee
where $\delta\ll1$ is an IR regulator expressing the distance from the boundary. 
See appendix \ref{app:massshift} for a derivation of \eqref{eq:massshiftresult}. Plugging in \eqref{eq:qufin}, setting $d=3$, and focusing on the logarithm part we get
\begin{equation}
\left.Q^{\partial}_{\mathrm{QU}}(P_1,P_2)\right|_\text{log}=-\frac{1}{g^2}  \sigma\frac{27}{2 \pi^6 (-2 P_1\cdot P_2)^4}\,.
\end{equation}

\subsubsection{Triple diagram}

The triple diagram reads
\begin{equation}
Q_{\text{TR}}(x_1,x_2)=g^2 \sigma\intads dx\, dy\ \Pi^{\lambda}_{\ \lambda'}(x,y) W_{\la \mu_1 \mu_2}(x,x_1,x_2)W^{\la'\mu_1\mu_2}(y,x_1,x_2)\,,
\label{eq:qutr}
\end{equation}
where we have introduced 
 \begin{equation}\label{eq:Wqutr}
 \begin{aligned}
     W_{\la\mu_1\mu_2}&(x,x_1,x_2)=2 \Pi_{\mu_1 \lambda}(x_1,x)\overset{\leftrightarrow}{\nabla}{}^\nu_x\Pi_{\nu \mu_2}(x,x_2)\\
     & -\Pi_{\mu_1 \nu}(x_1,x)\overset{\leftrightarrow}{\nabla}{}^x_{\lambda}\Pi^\nu_{\ \mu_2}(x,x_2)-\Pi_{\mu_1 \nu}(x_1,x)\overset{\leftrightarrow}{\nabla}{}^{\nu}_x\Pi_{\lambda \mu_2}(x,x_2)\,,
     \end{aligned}
 \end{equation}
 with $F\overset{\leftrightarrow}{\nabla}_\mu G=F(\nabla_\mu G)-(\nabla_\mu F)G$.\footnote{To obtain this expression for $W$ we have performed an integration by parts to get rid of the term in the triple vertex with a derivative acting on the external propagator. As we discuss in appendix \ref{app:intparts}, this does not give rise to boundary terms.}
Uplifting to embedding space and taking the boundary limit as in \eqref{eq:boundlimD}, we get 
\begin{equation}
 \begin{aligned}
Q^\partial_{\text{TR}}(P_1,P_2)&=\frac{1}{g^2} \sigma \int dX\, dY\ \Pi^C_{\ C '}(X,Y)W_{CA_1A_2}(X,P_1,P_2)W^{C'A_1A_2}(Y,P_1,P_2)\,,
     \end{aligned}
     \label{eq:qutrup}
 \end{equation}
 with
 \begin{equation}
     \begin{aligned}
         W_{CA_1A_2}&(X,P_1,P_2)=2  K_{A _1 C }(P_1,X)\overset{\leftrightarrow}\nabla{}^B_XK_{B  A _2}(X,P_2)\\&- K_{A _1 B }(P_1,X)\overset{\leftrightarrow}\nabla{}^X_CK^B_{\ A _2}(X,P_2)- K_{A _1 B }(P_1,X)\overset{\leftrightarrow}\nabla{}^B_XK_{C  A _2}(X,P_2)\,.
     \end{aligned}
 \end{equation}
We plug the expression for the bulk-to-boundary in \eqref{eq:boundprop} and the bulk-to-bulk gauge propagator in FY gauge in \eqref{subYennie}, getting in this way a linear combination of integrals in the form
\begin{equation}
   \mathcal{K}=\int dX \, dY (-2P_1\cdot X)^{-\Delta_1} (-2P_1\cdot Y)^{-\Delta_3} (-2P_2\cdot X)^{-\Delta_2} (-2P_2\cdot Y)^{-\Delta_4} f\left(u\left(X, Y\right)\right)\,.
\end{equation}
To solve this integral, we express $f(u(X,Y))$ as a function of the variable 
\begin{equation}\label{eq:zetaDef}
    \zeta= \frac{1}{1+u}=\frac{2 z w}{w^2+z^2+(x-y)^2}\,,
\end{equation}
and then expand it in powers of $\zeta$,
\begin{equation}
    f(\zeta)=\sum_{k=0}^\infty a_k \zeta^{\Delta_k}\,,
\end{equation}
with $\Delta_k=\Delta_0+k$. The result can be written as \be
\mathcal{K}=\sum_{k=0}^\infty a_k \, \mathcal{K}^{(k)}\,,
\ee
with 
\begin{align}
        \mathcal{K}^{(k)} =\sum_{m=0}^\infty \frac{
   \pi^\frac{d}{2}2^{\Di_k-1} \G\left(
    \frac{\Di_{1234}-d}{2}\right) \G\left(
    \frac{\Di_{34k}-d}{2}\right) \G\left(
    \frac{\Di_{4k,3}+2m}{2}\right) \G\left(
    \frac{\Di_{3k,4} }{2}\right) \G(m +\Di_3)}{ m!\ \G\left(\Di_1\right) \G\left(\Di_3\right) \G\left(\Di_k\right) \G\left(
    \frac{\Di_{124k}+\Di_4+2m-d}{2}\right)(-2 P_1\cdot P_2)^\frac{\Di_{12,k}-2m}{2}}  
      \mathcal{I}_{\tilde{\Delta}^{(k,m)}}\,.
 \label{eq:I4k}
      \end{align}
In \eqref{eq:I4k} $\mathcal{I}_\Delta$ is defined in \eqref{eq:massshift}, \eqref{eq:massshiftresult}, we introduced the notation 
\be
\Delta_{i_1i_2\dots,j_1j_2\dots}=(\Di_{i_1}+\Di_{i_2}+\dots)-(\Di_{j_1}+\Di_{j_2}+\dots)\,,
\label{eq:notDelta}
\ee
and we used that all the integrals that contribute to the triple diagram satisfy $\Di_{13}= \Di_{24}$, which gives
\be
\tilde{\Delta}^{(k,m)}=\frac{\Di_{34k} +2m}{2}\,.
\ee
We refer to appendix \ref{app:usefulint} for a derivation of the result \eqref{eq:I4k}. The sum over $m$ in \eqref{eq:I4k} can be performed analytically and gives rise to a linear combination of generalized hypergeometric functions depending on $k$. We numerically sum over $k$ setting $d=3$.\footnote{Actually, we have to set $d=3+\epsilon$, with $\epsilon\ll 1$, in order to avoid spurious poles in $1/(d-3)$ which appear in intermediate steps but cancel in the total sum.} The sum is convergent, but with a rather slow rate, so we adopt Pad\'e approximants to improve on the final accuracy.
The final result with 400 terms and [200/200] Pad\'e approximate is
\begin{equation}
\left.Q^{\partial}_{\mathrm{TR}}(P_1,P_2)\right|_\text{log}\approx \frac{1}{g^2} \sigma\frac{0.01976}{(-2 P_1\cdot P_2)^4}\,.
\end{equation}
Summing all the terms together and downlifting to Poincaré coordinates we get
\begin{equation}
\left.\langle J^{a_1}J^{b_1}(x_1) J^{a_2}J^{b_2}(x_2)\rangle^1\right|_\text{log}=\left.Q^{(1)\partial}(x_1,x_2)\right|_\text{log}\approx \frac{1}{g^2}\sigma \frac{0.005721}{x_{12}^8}\,.
\label{eq:JJ1}
\end{equation}
\subsection{Projecting onto irreducible representations}
As explained at the beginning of this section, to obtain primary operators we need to project $J^aJ^b$ onto irreducible representations of $SU(n_c)$. 
{According to \eqref{eq:rappsymm}, a symmetric tensor $T^{ab}$ decomposes as follows in terms of irreducible representations of $SU(n_c)$: 
\be
T^{ab} = \sum_{i=S,A,1,3} (P_i T)^{ab}\,,
\ee
where $P_i$ are the projectors to the appropriate representations, $S$ denoting the singlet representation.
The projectors are determined by imposing $P_i P_j = \di_{ij}P_j$, $\sum_i P_i = 1$ and ${\rm tr}\, P_i = {\rm dim}\; R_i$.
The singlet projector $P_S$ is easily found to be}
\begin{equation}
(P_S T)^{ab}=\frac{\delta_{cd}T^{cd}}{n_c^2-1}\delta^{ab}\,.
\end{equation}
The two-point function of the singlet operator
$JJ_S^{ab}=(P_SJJ)^{ab}$ is obtained by acting with this projector on eq.\eqref{eq:JJ} and eq.\eqref{eq:JJ1} as follows
\begin{equation}
    \langle JJ_S^{a_1b_1}(x_1) JJ_S^{a_2b_2}(x_2)\rangle=\delta^{a_1 b_1}\delta^{a_2 b_2}\left(\langle JJ_S(x_1) JJ_S(x_2)\rangle^0+ \langle JJ_S(x_1) JJ_S(x_2)\rangle^1\right)\,,
\end{equation}
with
 \begin{equation}
 \langle JJ_S(x_1) JJ_S(x_2)\rangle^0 =\frac{1}{g^4}     \frac{24}{\pi^4(n_c^2-1)} \frac{1}{x_{12}^8}\,,
 \end{equation}
\begin{equation}
\left.\langle JJ_S(x_1) JJ_S(x_2)\rangle^1\right|_\text{log}\approx \frac{1}{g^2} \frac{2n_c}{n_c^2-1}\frac{0.005721}{x_{12}^8}\,.
\end{equation}
Plugging in \eqref{eq:gammaprim}, we get that the anomalous dimension of the singlet operator is
\begin{equation}
\gamma_{JJ_S}(g^2) \approx -0.04644\, n_c\, g^2 + {\cal O}(g^4)\,.
\end{equation}
This result matches with \eqref{eq:gammaJJdisp}, providing a non-trivial check of our computation. This correspondence allows to identify the 0.005721 in \eqref{eq:JJ1} as the analytic result $11/(2 \pi^6)$, which we replace from now on. 

Consider now the adjoint representation. The projector reads
\begin{equation}\label{eq:projAdj}
(P_A T)^{ab}=\frac{n_c}{n_c^2-4}d^{abe}d_{cde}T^{cd}\,,
\end{equation}
$d_{abc}=2\,{\rm tr}\,[\{t_a,t_b\}t_c]$.
Applying \eqref{eq:projAdj} to eq.\eqref{eq:JJ} and eq.\eqref{eq:JJ1}, we  obtain the following expression for the two-point function of $JJ_{A}^{ab}=(P_AJJ)^{ab}$,
\begin{equation}
\begin{aligned}
    \langle JJ_{A}^{a_1b_1}(x_1) JJ_{A}^{a_2b_2}(x_2)\rangle&=d^{a_1 b_1 c} d^{a_2 b_2}_{\quad \ c}\left(\langle JJ_{A}(x_1) JJ_{A}(x_2)\rangle^0+ \langle JJ_{A}(x_1) JJ_{A}(x_2)\rangle^1\right)\,,
    \end{aligned}
\end{equation}
with
 \begin{equation}
 \langle JJ_{A}(x_1) JJ_{A}(x_2)\rangle^0 = \frac{1}{g^4}    \frac{24 n_c}{\pi^4(n_c^2-4)} \frac{1}{x_{12}^8}\,,
 \end{equation}
\begin{equation}
\left.\langle JJ_{A}(x) JJ_{A}(y)\rangle^1\right|_\text{log}= \frac{1}{g^2}\frac{11n_c^2}{2\pi^6(n_c^2-4)}\frac{1}{x_{12}^8}\,. 
\end{equation}
To compute these expressions, we use known relations involving the structure constants $f_{abc}$ and the symmetric tensors $d_{abc}$, see e.g.\cite{Haber:2019sgz}. 
By taking the ratio of the last two equations we find the anomalous dimension for the adjoint representation to be
\begin{empheq}[box=\mybluebox]{align}\label{eq:gammaJJadjoint}
\gamma_{JJ_{A}}(g^2)=- \frac{11}{48\pi^2}\, {n_c} g^2 + {\cal O}(g^4)\,.
\end{empheq}
For completeness, let us also consider the remaining representations $R_1$ and $R_3$. 
The projectors read
\begin{equation}
\begin{aligned}
        (P_1T)^{ab}&=\frac{1}{4}\left(\delta^{a}_{\ c}\delta^{b}_{\ d}+\delta^{a}_{\ d}\delta^{b}_{\ c}-f^{e a}_{\ \ c}f^b_{\ de}-f^{ea}_{\ \ d}f^b_{\ ce}+\frac{n_c}{n_c+2}d^{abe}d_{cde}+2\frac{\delta^{ab}\delta_{cd}}{n_c+1}\right)T^{cd}\,,\\
        (P_3T)^{ab}&=\frac{1}{4}\left(\delta^{a}_{\ c}\delta^{b}_{\ d}+\delta^{a}_{\ d}\delta^{b}_{\ c}+f^{ea}_{\ \ c}f^{b}_{\ de}+f^{e a}_{\ \ d}f^b_{\ ce}-\frac{n_c}{n_c-2}d^{abe}d_{cde}-2\frac{\delta^{ab}\delta_{cd}}{n_c-1}\right)T^{cd}\,.\\
\end{aligned}
\end{equation}
Proceeding as before, we get
\begin{equation}\label{eq:gammaR1R3}
    \begin{aligned}
    \gamma_{JJ_{1}}(g^2)&=\frac{11}{24\pi^2}\,  g^2 + {\cal O}(g^4)\,,\\
    \gamma_{JJ_{3}}(g^2)&=- \frac{11}{24\pi^2}\,  g^2 + {\cal O}(g^4)\,.
    \end{aligned}
\end{equation}
The results \eqref{eq:gammaJJadjoint} and \eqref{eq:gammaR1R3} apply for any $n_c >3$. For $n_c=3$ the operator $JJ_{3}$ does not exist. 
For $n_c=2$ also $JJ_A$ does not exist, the lightest adjoint scalar having in this case  classical dimension $\Delta=7$, see appendix \ref{app:characters}.

\section{Current two-point function}
\label{sec:cj}
In this section we compute the current two-point function \eqref{eq:JJ2pt} at the next-to-leading order.
Since there is no anomalous dimension for a conserved current, the correction amounts to rescaling of the correlator. In embedding space, this reads
\be
K(P_1,P_2)=\langle J_{A_1}(P_1) J_{A_2}(P_2)\rangle  =C_J\frac{{\cal P}_{A_1A_2}(P_1,P_2)}{(-2P_1\cdot P_2)^{d-1}}\,,
\label{eq:JJemb}
 \ee
 where $\calP_{A_1A_2}(P_1,P_2)$ is the boundary limit of \eqref{eq:projc},
and we have introduced the corrected normalization constant
\begin{equation}
C_J=\frac{C^0_J}{g^2}\Big(1+C^1_J g^2+\mathcal{O}(g^4)\Big)\,.
\end{equation}
We determine $C_J^1$ by computing the one-loop corrections to the $D$ bulk-to-bulk gauge propagator in ambient configuration space.
After that, we uplift the result in embedding space and take the boundary limit. We then evaluate the necessary integrals
and finally extract the value of $C_J^1$. We work in the FY gauge. 
 
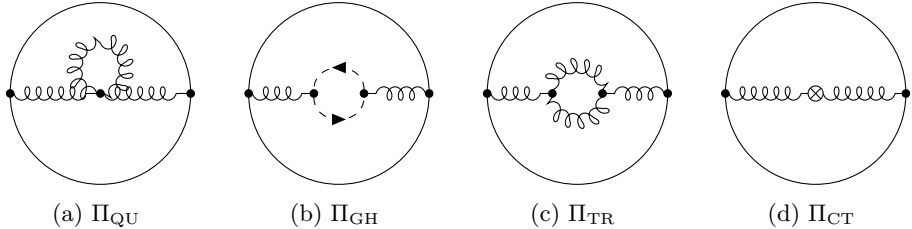
\begin{figure}[t!]
  \centering 
  \begin{subfigure}[c]{0.15\textwidth}
  \begin{minipage}{4cm}
\begin{tikzpicture}
  \begin{feynman}[inline=(d.base)]
  \vertex (a) at (0,0)[dot] {\small{}};
   \vertex (i1) at (-1.2,0)[dot]{\small{ }} ;
    \vertex (b) [dot] at (1.2,0){\small};
  \vertex (c) at (0,2/3);
    \draw (0,0) circle (1.2);

    \diagram* {
      (i1) -- [gluon] (a),
      (a) -- [gluon, half left] (c),
      (c) -- [gluon, half left] (a),
      (a) -- [gluon] (b),
    };
  \end{feynman}
\end{tikzpicture}  
\end{minipage}
        \caption{$\Pi_{\rm QU}$}
    \end{subfigure}\qquad 
            \begin{subfigure}[c]{0.15\textwidth}
\begin{minipage}{4cm}

                    \begin{tikzpicture}
  \begin{feynman}[inline=(a.base)]
  \draw (0,0) circle (1.2);
  \vertex (a) at (-1/3,0) [dot]{\small{ }};
    \vertex (i1) at (-1.2,0)[dot]{\small{ }} ;
    \vertex (b) at (1/3,0)[dot]{\small{ }};
    \vertex  (o2) at (1.2,0)[dot] {\small{  }};

    \diagram* {
      (i1) -- [gluon] (a),
      (a) -- [charged scalar, half right] (b) -- [charged scalar
, half right] (a),
      (o2) -- [gluon] (b),
    };
  \end{feynman}
\end{tikzpicture} 
        \end{minipage}
        \caption{$\Pi_{\rm GH}$}
        \end{subfigure}\qquad \begin{subfigure}[c]{0.15\textwidth}
    \begin{minipage}
        {4cm}
        \begin{tikzpicture}
  \begin{feynman}[inline=(a.base)]
\draw (0,0) circle (1.2);
  \vertex (a) at (-1/3,0)[dot]{\small{}};
    \vertex (i1) at (-1.2,0)[dot]{\small{ }} ;
    \vertex (b) [dot] at (1/3,0){\small};
    \vertex [dot] (o2) at (1.2,0) {\small{  }};

    \diagram* {
      (i1) -- [gluon] (a),
      (a) -- [gluon, half left] (b) -- [gluon
, half left] (a),
      (o2) -- [gluon] (b),
    };
  \end{feynman}
\end{tikzpicture}  
\end{minipage}
        \caption{$\Pi_{\rm TR}$}
        \end{subfigure}\qquad
            \begin{subfigure}[c]{0.15\textwidth}
                \begin{tikzpicture}
  \begin{feynman}[inline=(a.base)]
  \draw (0,0) circle(1.2);
  \vertex (a) at (0,0) [crossed dot]{};
    \vertex [dot](i1) at(-1.2,0){\small{ }} ;
    \vertex [dot] (b) at (1.2,0){};

    \diagram* {
      (i1)--[gluon](a)--[gluon](b)
    };
  \end{feynman}
\end{tikzpicture} 
     \caption{$\Pi_{\rm CT}$}
    \end{subfigure}
\caption{One-loop corrections to the current $J$ two-point function: the quartic (a), ghost (b), triple (c), and counterterm (d) diagram. }
\label{fig:su2}
\end{figure}
\subsection{Computation of the diagrams: external points in the bulk}
\label{sec:diagrams}
The leading perturbative corrections of the vector two-point function is given by the one-loop diagrams depicted in figure \ref{fig:su2}. They read schematically
\begin{equation}g^2 {\Pi}_{\text{tot}}^{(1)}(x_1,x_2)=\langle A(x_1) A(x_2) \rangle = g^2\Pi^{(1)}(x_1,x_2)+g^2\Pi_{\mathrm{CT}}(x_1,x_2)\,,
    \label{eq:G2tot}\end{equation}
    where $x_1$ and $x_2$ are points in the bulk that we send to the boundary at the end of the calculation. The one-loop contribution $\Pi^{(1)}$ is the sum of three diagrams:
    \be
  \Pi^{(1)}  =\Pi_{\mathrm{QU}}+\Pi_{\mathrm{GH}}+\Pi_{\mathrm{TR}}\,,
    \ee
which we call quartic, ghost, and triple diagram. $\Pi_{\mathrm{CT}}$ denotes the one-loop counterterm contribution. 
Since the color structure of each contribution is diagonal in the color indices, 
$(\Pi_{\bullet})^{a_1 a_2}_{\mu_1 \mu_2}(x_1,x_2) = \delta^{a_1 a_2}(\Pi_{\bullet})_{\mu_1 \mu_2}(x_1,x_2)$,
we omit the color indices in what follows.

\subsubsection{Quartic diagram}

The quartic diagram $\Pi_{\mathrm{QU}}$ is the simplest as it involves integrating only over a bulk coordinate $x$. It reads
\begin{equation}
    (\Pi_{\mathrm{QU}})_{\mu_1 \mu_2}(x_1,x_2) = -n_c \, g^2\!
 \int \! dx \,\Pi_{\mu_1\lambda}(x_1,x)\Pi_{\mu_2\nu}(x_2,x)\Pi_{\rho\sigma}(x,x)(g^{\lambda\nu} g^{\rho\sigma}-g^{\lambda\rho} g^{\nu\sigma})
\,. 
\end{equation}
We can evaluate the tadpole $\Pi_{\rho\sigma}(x,x)$ by expanding the gauge propagator for small values of the chordal distance. In dimensional regularization, the only surviving term in the expansion is the constant one, but remarkably this vanishes when we select the FY gauge.
This is not in contrast with the results of flat space, where the diagram is zero regardless of the value of $\xi$, because if we reintroduce the dependence on the AdS radius $L$, we see that the diagram is of order $\mathcal{O}(1/L^2)$, and therefore its contribution vanishes in the flat space limit.

\subsubsection{Ghost diagram}
The ghost contribution $\Pi_\text{GH}$ reads
\begin{equation}
\begin{aligned}
(\Pi_\text{GH})_{i_1 i_2}&(x_1,x_2)=n_c g^2\intads dx\, dy \ \Pi_{i_1}^{\lambda}(x_1,x)  \Pi_{i_2}^{\lambda^{\prime}}(x_2,y)(\nabla_\lambda G_\text{GH}\nabla_{\lambda^\prime} G_\text{GH})\,.
\label{eq:ghostdiag}
\end{aligned}
\end{equation}
Replacing the expression of the ghost propagator \eqref{eq:GHprop} and using the basis of eq. \eqref{eq:bitensor}, we obtain 
\begin{equation}\label{eq:PiGh1}
\begin{aligned}
\nabla_\lambda G_\text{GH}\nabla_{\lambda^\prime} G_\text{GH} = f_{\text{GH1}}(u)g_{\lambda \lambda'}+f_{\text{GH2}}(u)n_\lambda n_{\lambda'}\,,
\end{aligned}
\end{equation}
where 
\begin{equation}
\begin{aligned}
&f_{\text{GH1}}(u)=0\,,\quad f_{\text{GH2}}(u)=\frac{\Gamma\left(\frac{d+1}{2}\right)^2}{4 \pi^{d+1} (u (2 + u))^{d} }\,.
\end{aligned}
\end{equation}
Note that despite the ghost propagator involves hypergeometric functions, the combination \eqref{eq:PiGh1} gives rise to meromorphic functions
of the same kind as those appearing in the gauge propagator \eqref{subYennie} in the FY gauge. 

\subsubsection{Triple diagram}

We now turn to the computation of the triple gauge diagram $\Pi_\text{TR}$, which is the most involved.  
Considering all Wick contractions among the two triple couplings, integration by parts, and after some algebra, we get
\begin{align}
(\Pi_\text{TR})&_{i_1 i_2}(x_1,x_2)=n_c  g^2\!\intads dx \, dy \ \Pi_{i_1 \lambda}(x_1,x)  \ \Pi_{i_2 \lambda^{\prime}}(x_2,y)\bigg(4\left( \Pi^{\nu \nu^{\prime}} \nabla_\nu \nabla_{\nu'}  \Pi^{\lambda \lambda^{\prime}}-\nabla_\nu  \Pi^{\lambda \nu^{\prime}} \nabla_{\nu'}  \Pi^{\nu \lambda^{\prime}}\right)\nn \\
&-2\left( \Pi_{\nu \nu^{\prime}} \nabla^\nu \nabla^{\lambda^{\prime}}  \Pi^{\lambda \nu^{\prime}}-\nabla^\nu  \Pi^{\lambda \nu^{\prime}} \nabla^{\lambda^{\prime}}  \Pi_{\nu \nu^{\prime}}\right)-2\left( \Pi_{\nu \nu^{\prime}} \nabla^\lambda \nabla^{\nu^{\prime}}  \Pi^{\nu \lambda^{\prime}}-\nabla^\lambda  \Pi_{\nu \nu^{\prime}} \nabla^{\nu^{\prime}}  \Pi^{\nu \lambda^{\prime}}\right)\nn \\
&+\left( \Pi_{\nu \nu^{\prime}} \nabla^\lambda \nabla^{\lambda^{\prime}}  \Pi^{\nu \nu^{\prime}}-\nabla^\lambda  \Pi_{\nu \nu^{\prime}} \nabla^{\lambda^{\prime} } \Pi^{\nu \nu^{\prime}}\right)+\left( \Pi^{\lambda \lambda^{\prime}} \nabla_\nu \nabla_{\nu^{\prime}}  \Pi^{\nu \nu^{\prime}}-\nabla_\nu  \Pi^{\nu \lambda^{\prime}} \nabla_{\nu^{\prime}}  \Pi^{\lambda \nu^{\prime}}\right)  \label{eq:tripleabs} \\
&-2\left(\nabla_{\nu^{\prime}} \nabla_\nu  \Pi^{\nu \lambda^{\prime}}  \Pi^{\lambda \nu^{\prime}}-\nabla_{\nu^{\prime}}  \Pi^{\lambda \lambda^{\prime}} \nabla_\nu  \Pi^{\nu \nu^{\prime}}\right)-2\left( \Pi^{\nu \lambda^{\prime}} \nabla_\nu \nabla_{\nu^{\prime}}  \Pi^{\lambda \nu^{\prime}}-\nabla^\lambda  \Pi^{\nu \lambda^{\prime} }\nabla^{\nu^{\prime}}  \Pi_{\nu \nu^{\prime}}\right)\nn \\
&+\left(\nabla^{\lambda^{\prime}} \nabla^\nu  \Pi_{\nu \nu^{\prime}}  \Pi^{\lambda \nu^{\prime}}-\nabla^{\lambda^{\prime}}  \Pi^{\lambda \nu^{\prime}} \nabla^\nu  \Pi_{\nu \nu^{\prime}}\right)+\left( \Pi^{\nu \lambda^{\prime}} \nabla^\lambda \nabla^{\nu^{\prime}}  \Pi_{\nu \nu^{\prime}}-\nabla^\lambda  \Pi^{\nu \lambda^{\prime} }\nabla^{\nu^{\prime}}  \Pi_{\nu \nu^{\prime}}\right)\bigg)\,. \nn
\end{align}
Plugging the expression \eqref{eq:propintr} for the gauge propagator in the basis of \cite{Allen1986} and using eq.~\eqref{eq:useful}, we can rewrite this term in the form 
\begin{equation}\label{eq:tripleExp} 
\begin{aligned}
(\Pi_\text{TR})_{i_1 i_2}&(x_1,x_2)=n_c g^2\intads dx\, dy \ \Pi_{i_1 \lambda}(x_1,x)  \Pi_{i_2 \lambda^{\prime}}(x_2,y)f_\text{TR}^{\ \lambda \lambda'}(x,y)\,,
\end{aligned}
\end{equation}
with
\begin{equation}\label{eq:fTREXP}
\begin{aligned} 
    &f_\text{TR}^{  \ \lambda \lambda'}(x,y)=f_{\text{TR1}}(u)g^{\lambda \lambda'}+f_{\text{TR2}}(u)n^\lambda n^{\lambda'},\\
&f_{\text{TR1}}(u)=\frac{(d-4) (d-1) \Gamma\left(\frac{
  d+1}{2}\right)^2(1 + u)}{4 (d-2)^2  \pi^{d+1}(u (2 + u))^d} \,,\\
& f_{\text{TR2}}(u)=\frac{ \Gamma\left(\frac{
  d+1}{2}\right)^2\Big(4 (1 + u)+ d \big(d-6 + u (d -5)\big) \Big)}{4 (d-2)^2  \pi^{d+1}(u (2 + u))^{d}} \,.
  \end{aligned}
\end{equation}
In deriving \eqref{eq:tripleExp} from \eqref{eq:tripleabs} one has to pay attention to possible contact terms, which can arise since 
eq. \eqref{eq:tripleabs} contains propagators inside the loop that are derived twice. A way to take care of this is by replacing the equation of motion for the gauge propagator \eqref{eq:gpAdS3}
in \eqref{eq:tripleExp}. However, since contact terms produce tadpoles of the gauge propagator, these contributions vanish in the FY gauge, as it happens in the quartic diagram. 

\subsubsection{Total contribution}

The sum of the ghost and triple diagram contributions $\Pi^{(1)}=\Pi_\text{GH}+\Pi_\text{TR}$ can be rewritten, interestingly enough, in terms of gauge propagators only in the following form,
\begin{equation}
\begin{aligned}
(\Pi^{(1)})_{i_1 i_2}(x_1,x_2)=\Big(\frac 4d -1\Big)n_c\  g^2 
\!\intads dx\, dy \ \nabla_\nu\Pi_{i_1 \lambda}(x_1,x)  \nabla_{\nu^\prime} \Pi_{i_2 \lambda^{\prime}}(x_2,y)( \Pi^{\nu \nu^{\prime}} \Pi^{\lambda \lambda^{\prime}}-\Pi^{\lambda \nu^{\prime}} \Pi^{\nu \lambda^{\prime}}) \,.
\label{eq:eqfinal}
\end{aligned}
\end{equation}
The expression \eqref{eq:eqfinal} has been conveniently written in a way in which the derivatives act on each of the two external legs.
In this way, we avoid double derivatives acting on external propagators, which may be generated after integration over one of the internal points. The resulting contact terms, 
contrary to those discussed below eq. \eqref{eq:fTREXP}, in general would not vanish.

\subsection{Counterterm} 

The counterterm contribution reads
\begin{align}
(\Pi_{\mathrm{CT}})_{\mu_1 \mu_2}(x_1,x_2) = -g^2&\bigg(
\delta_T \intads dx\ \Big(\nabla^\nu\Pi_{\mu_1\lambda}(x_1,x)\nabla^{\lambda}\Pi_{\mu_2\nu}(x_2,x)\Big) \nn \\
&\quad+ \delta_L \intads dx\  \Big(\frac{1}{\xi}\nabla^\nu \Pi_{\mu_1 \nu}(x_1,x)\nabla^\lambda\Pi_{\mu_2\lambda}(x_2,x)\Big)\bigg)
\,, 
\label{eq:diagram1}
\end{align}
where the coefficients $\delta_T$ and $\delta_L$ are a flat space result. In modified  minimal subtraction ($\overline{\text{MS}}$) scheme we have (see e.g. \cite{Schwartz:2014sze})
\be\label{eq:CTdeltaT}
\delta_T = \frac{ n_c}{32\pi^2}\left(\frac{10}{3}+(1-\xi)\right)\left(\frac{2}{\epsilon}-\gamma_E+\log(4\pi)\right)\ +\mathcal{O}(g^4)\,,\qquad \delta_L = 0\,.
\ee

\subsection{Final result}

We uplift in embedding space \eqref{eq:eqfinal} and \eqref{eq:diagram1} and take the boundary limit of the points $X_{1,2}$. This gives
\begin{align}
(K^{(1)})_{A_1 A_2}(P_1,P_2) & = \Big(\frac 4d -1\Big) n_c \int d X\, dY\ \nabla_B K_{A_1 A }(P_1,X)  \nabla_{B^\prime} K_{A_2 A ^{\prime}}(P_2,Y)\nn \\
& \hspace{3cm} \times ( \Pi^{B B^{\prime}} \Pi^{A  A ^{\prime}}-\Pi^{A  B^{\prime}} \Pi^{B A ^{\prime}}) \,, \label{eq:sumfinal} \\
 (K_{\mathrm{CT}})_{A_1 A_2}(P_1,P_2) & = -\delta_T\! \int \! dX \,\nabla^C K_{A_1B}(P_1,X)\nabla^{B}K_{A_2 C}(P_2,X)
\,.  \label{eq:count}
\end{align}
The loop and counterterm contributions can be written in terms of scalar contributions $K^{(1)}$ and $K_{\mathrm{CT}}$ as follows,
\begin{align}
\begin{split}
(K^{(1)})_{A_1 A_2}(P_1,P_2) & \equiv  {\cal P}_{A_1A_2}(P_1,P_2) K^{(1)}\,,  \\
(K_{\mathrm{CT}})_{A_1 A_2}(P_1,P_2) & \equiv {\cal P}_{A_1A_2}(P_1,P_2) K_{\mathrm{CT}}\,.
\label{eq:ctbis0}
\end{split}
\end{align} 
We evaluate the integrals appearing in \eqref{eq:sumfinal} and \eqref{eq:count}, starting from the simpler counterterm contribution.
We insert the expression for the $D$ bulk-to-boundary propagator \eqref{eq:boundprop} and use Lorentz invariance to reduce \eqref{eq:count} to scalar integrals only.
The resulting expression can be written in the form \eqref{eq:ctbis0} with 
\begin{equation}
\label{eq:ctbis}
\begin{aligned}
K_{\mathrm{CT}} &= -
\delta_T \frac{\Gamma(d)^2}{ \Gamma(\frac{d}{2})^2\pi^d d}\int \! dX  \frac{(d-1)(P_1\cdot P_2)+(d-2)(P_1\cdot X)(P_2\cdot X) }{\left(-2P_1 \cdot X\right)^d\left(-2P_2 \cdot X\right)^d}\\
&=  \delta_T \frac{\Gamma\left(\frac{d+1}{2}\right)(d-2)}{ 2\pi^\frac{d+1}{2} d^2(-P_1\cdot P_2)^{d-1}}\,,
\end{aligned}
\end{equation}
where we used \eqref{eq:massshiftresult} in the last step.
Note that the log terms appearing in \eqref{eq:massshiftresult} cancel in \eqref{eq:ctbis}, as expected, since the currents $J_i^a$ are conserved and cannot acquire an anomalous dimension.

We consider now $\Pi^{(1)}$ in \eqref{eq:sumfinal}.
First, we use the expression of the bulk-to-boundary and bulk-to-bulk gauge propagators and act with an inversion transformation (see the beginning of section \ref{app:usefulint} in the appendix for the detailed form of the transformation). After some algebra, the integral can be written as in \eqref{eq:ctbis0}, namely
\begin{equation}
    K^{(1)}= \int\! dX  \frac{(d-1)(P_1\cdot P_2)+(d-2)(P_1\cdot X)(P_2\cdot X) }{\left(-2P_1 \cdot X\right)^d\left(-2P_2 \cdot X\right)} I_Y(z),
    \label{eq:KTOT}
\end{equation}
with 
\begin{equation}\label{eq:IY}
I_Y(z)=\kappa_d \int_0^\infty\!{dw}\int_0^\infty\!{ d y} \ \frac{w^{2d-3}y^{d+1}z^{2d-1}}{(y^2+(w+z)^2)^d{(y^2+(w-z)^2)^d}}\,,
\end{equation}
and \begin{equation}
\kappa_d=\Big(\frac 4d -1\Big) n_c \frac{2^{4d-5}(d-1) \Gamma(\frac{d+1}{2})^4\Omega_{d-1}}{d^2(d-2)^2 \pi^{2d+2}},
\end{equation}
where $\Omega_{d-1}= 2\pi^{\frac{d}{2}}/\Gamma(\frac{d}{2})$ is the volume of the $(d-1)-$dimensional sphere with unit radius. 
It is useful to introduce two Schwinger parameters $t_1$ and $t_2$ and rewrite $I_Y(z)$ as 
\begin{align}
\begin{split}
I_Y(z)=\frac{\kappa_d}{\Gamma(d)^2} \int_0^\infty \!\!\! d t_1  \int_0^\infty\! \! \!d t_2\int_0^\infty \!\!\! {dw}\! \int_0^\infty\! \! \! {d y} \ w^{2d-3}y^{d+1}&z^{2d-1}{t_1^{d-1}t_2^{d-1}} \\ & \times {e}^{-(y^2+(w+z)^2)t_1}{e}^{-(y^2+(w-z)^2)t_2}\,.
\end{split}
\end{align}
The integral over $y$ and $w$ can be computed analytically and we are left with an integral over $t_1$ and $t_2$. Following \cite{Bertan:2018afl}, we perform a change of variables $t_1=u s, \ t_2=(1-u)s$.
The integral over $s$ can be computed analytically and we are left with the integral over $u$, which is UV divergent. 
This last integral is computed using the following trick \cite{DeCesare:2022obt}: we isolate the divergences by expanding around coincident points ($u=0$) up to a sufficient order so that the remaining part is finite and can be safely computed. The divergent terms are regulated by using
\begin{equation}
    \int_0^1du\ u^{a-1}=\frac{1}{a}\, ,
    \label{eq:maintool}
\end{equation}
which is valid for $a>0$, but is extendable to any $d$-dependent $a$ by analytic continuation in $d$. We then sum both contributions to obtain
\begin{equation}\label{eq:IYzExp}
I_Y(z)=z^{d-1}n_c 2^{d+1} \Big(\frac 4d -1\Big) \left(\frac{1}{12\pi^6\epsilon}+\frac{-1+ 9 \gamma_E + 9 \log\pi}{72 \pi^6}\right)\,.
\end{equation}
The $z^{d-1}$ term appearing in \eqref{eq:IYzExp} is such that $(-2P_2\cdot X)$ in \eqref{eq:KTOT} turns into $(-2P_2\cdot X)^d$ thanks to \eqref{eq:p2inv}.
The $dX$ integral in \eqref{eq:KTOT} can then be evaluated using again \eqref{eq:massshift} and \eqref{eq:massshiftresult}. We get 
\be\label{eq:K1Exp}
K^{(1)} = \frac{\pi^\frac{d}{2}(d-4) (d-2)   \Gamma\left(\frac{d}{2}\right)}{ d^2  \Gamma(d)(-P_1\cdot P_2)^{d-1}}\left(\frac{1}{6\pi^{6}\epsilon}+\frac{-1 + 9\gamma_E + 9 \log\pi}{36\pi^{6}}\right)\,.
\ee
Again, the log terms in \eqref{eq:massshiftresult} cancel in \eqref{eq:K1Exp}, as expected.
We sum loop and counterterm contribution, and set $\xi=d/(d-2)$ in $\delta_T$ in \eqref{eq:CTdeltaT}. In this way we get
\begin{equation}
 K_{\text{tot}}^{(1)} = K_{\text{CT}}+K^{(1)} =  -n_c\frac{10 + 3 \gamma_E}{162 \pi^4} \frac{1}{(-2P_1\cdot P_2)^2}\,.
\end{equation}
The cancellation of the UV divergences, as expected from the renormalization of the theory, provides a sanity check of the computation.
Matching with eq. \eqref{eq:JJemb}, finally allows us to determine the correction to $C_J$,
\begin{empheq}[box=\mybluebox]{align}{
    C_J^1=-\frac{10 + 3 \gamma_E}{324 \pi^2}\ n_c\,.}
\label{eq:CJ1Fin}    
\end{empheq}
From this result we extract the estimate \eqref{eq:Decoup} for the Decoupling scenario. Note that the value of $C^1_J$ is renormalization scheme-dependent. The value \eqref{eq:CJ1Fin}
is in the $\overline{\rm MS}$ scheme.\footnote{Beyond the leading order computed in this paper, also the $g^2$ expansion of boundary scaling dimensions becomes scheme-dependent. One way to get rid of scheme-dependence is to eliminate $g^2$ and express the scaling dimensions, that are physical, as an expansion in the physical quantity $1/C_J$. Doing so, all the coefficients in the expansion are themselves physical. Given the function $\Delta_{\mathrm{tr}[JJ]}(1/C_J)$, the question of Marginality vs Decoupling becomes the question of whether $\Delta_{\mathrm{tr}[JJ]}(1/C_J) = 3$ for any positive $C_J$.}

\section{Conclusion}\label{sec:conc}

In this paper, 
inspired by \cite{Aharony:2012jf},
we have explored confinement in non-abelian gauge theories in AdS$_4$, from the perspective of the boundary CFT${}_3$. 
Among the three possibilities reviewed in the introduction, our results favor the Marginality one.

There are several open questions that would be important to address in future studies. 
The merging scenario implies the existence of a new theory $D^*$ which has the same global symmetry of the $D$ CFT. 
Finding possible candidates for $D^*$ is an important point that we did not address. In particular, it would be useful to see if  there exist candidates for $D^*$ 
in the vicinity of $g^2=0$, i.e. at weak coupling. While the possibility illustrated in figure \ref{fig:DDstarN} appears to be the most likely, more work is needed to firmly exclude more exotic possibilities. For instance it is in principle possible that the symmetry $G$ appears as an emergent symmetry in the $N$ bc, allowing $D$ and $N$ to annihilate, leaving some other boundary condition at strong coupling. Even in the scenario of figure \ref{fig:DDstarN}, an important question is whether the theory settles to the $N$ boundary condition after the merger, and if this is the case, whether this happens continuosly or discontinuosly. As pointed out in \cite{Copetti:2023sya}, anomalies in generalized symmetries can sometimes rigorously rule out the continuity between $N$ and $D$. It is also in principle possible that, for some reason, multiple scenarios occur simultaneously at the same value of $g^2_\text{crit}$.

Having reformulated confinement purely in terms of properties of a non-local boundary CFT, it would be extremely interesting to see if 
the conformal bootstrap \cite{Poland:2018epd, Rychkov:2023wsd} might be used to rigorously assess which scenarios are consistent. Recent progress in the study of 
four-point functions of non-abelian conserved currents in 3d \cite{He:2023ewx} make this direction feasible in the near future.

From a more theoretical point of view, it would be important to properly define what confinement means in AdS space.
As is well known, in flat space confinement is detected by the area law of large Wilson loops. Recently, the area law and confinement have been reformulated more sharply
as the phase in which (e.g. for $SU(N)$ gauge theories) the electric $\mathbb{Z}_N$ one-form symmetry is unbroken \cite{Gaiotto:2014kfa}. In AdS at finite $L$, there is no
intrinsic distinction between perimeter and area law, and hence it is not clear if one-form symmetries still characterize the possible phases. 
The space has however a boundary and perhaps a sharp characterization is provided by the boundary conditions. In fact, the one conjectured in our paper is one of these:
confinement in AdS is characterized by the absence of the $D$ bc. It would be important to understand how this definition is related to the usual one
in terms of one-form symmetries and to verify if other definitions are in principle possible. 

Finally, it would be interesting to explore possible applications of the general result we found for the scaling dimension of the displacement operator \eqref{eq:DisNoMargEx} and \eqref{eq:CoCTD3}.
Recently correlation functions involving the bulk stress tensor and boundary operators along bulk RG flows were studied in \cite{Meineri:2023mps, Lauria:2023uca}. These papers derived sum rules for the scaling dimension of the boundary operator, which can be applied in particular to the displacement operator. It would be interesting to compare the sum rules to the result for the scaling dimension of $\mathcal{D}$ presented here. Matching the two results, it should be possible to obtain a sum rule for the bulk beta function, e.g. to express the one-loop beta function in terms of a sum involving boundary CFT data. In the context of amplitudes in flat space, it has been shown that RG coefficients, including beta function coefficients, can be extracted from scattering data, see e.g. ref.s \cite{Caron-Huot:2016cwu, EliasMiro:2020tdv, Baratella:2020lzz, Baratella:2020dvw}. It would be interesting to explore further how boundary correlation functions in AdS encode bulk RG coefficients and the relation to the flat space results via the flat space limit.

\section*{Acknowledgments}
We thank Andrea Antinucci, Stéphane Bajeot, Davide Bason, Agnese Bissi, Dean Carmi, Christian Copetti, Simone Giombi, Victor Gorbenko, Ziming Ji, Shota Komatsu, Marco Meineri, Tomáš Procházka, and Veronica Sacchi for discussions and/or collaboration on related projects. FDC also expresses deep gratitude to Princeton University for hospitality during part of this work. Work partially supported by INFN Iniziativa Specifica ST\&FI. The work of RC is supported by the Israel Science Foundation, grant no. 1487/21, and by the MOST NSF/BSF Physics grant no. 2022726.

\appendix
\section{Mean field theory of $SU(n_c)$ adjoint currents}
\label{app:characters}
The spectrum of CFT operators on $\RR^d$, or equivalently of states on $S^{d-1}\times \RR$, can be encoded in a grand-canonical partition function on $S^{d-1}\times S^1$, with $S^1$ being a compact Euclidean thermal circle. In particular, we are interested in the spectrum of the mean-field theory of $SU(n_c)$ adjoint currents in $d=3$ dimensions. Following the approach of \cite{Sundborg:1999ue,Aharony:2003sx}, this can be determined from the single-particle partition function 
\begin{equation}
    z_{J,{R_A}}(q,x,y,r)=z_{J}(q,x)X_{R_A}(y,r)\,.
\end{equation}
Here,
\begin{equation}
    z_{J}(q,x)={\rm Tr}_{J}\left(q^\Delta x^j\right)=\chi_{(2,1)}(q,x)-\chi_{(3,0)}(q,x)\equiv \chi^{short}_{(2,1)}(q,x)\,,
\end{equation}
is the single-particle partition function of a $U(1)$ conserved current $J$ in $d=3$, with
\begin{equation}
    \chi_{(\Delta,\ell)}(q,x)=\frac{q^\Delta}{(1-q)(1-q x)(1-q/x)}\sum_{j=-\ell}^\ell x^j\,,\qquad q=e^{-\beta}\,,\ \ x=e^{\mu}\,,
\end{equation}
being the conformal characters associated with primary operators with scaling dimension $\Delta$ and $SO(3)\sim SU(2)$ spin $\ell$, for which we have turned on fugacities $q$ and $x$, respectively. Similarly,
\begin{equation}
    X_{R_A}(y,r)= y^{-r}\left(\sum_{p=0}^r y^p\right)^2 -1\,,
\end{equation}
is the character for the $SU(r+1)$ adjoint representation $R_A$, with a common fugacity $y$ for the diagonal Cartan generators. The spectrum of the mean-field theory of $SU(r+1)$ adjoint currents is encoded in the multi-particle partition function
\begin{equation}
    Z(q,x,y,\eta,r)=\exp\left(\sum_{k=1}^\infty \eta^k\frac{z_{J,R_A}(q^k,x^k,y^k,r)}{k}\right)\,,
\end{equation}
where we have also introduced a fugacity $\eta$ that keeps track of the number of currents entering each primary operator.
The partition function $Z$ can be systematically expanded in powers of $q,x,\eta$ in order to obtain the spectrum to arbitrary order. Up to scaling dimension $\Delta=7$, omitting to write the common $(q,x)$-dependence on all characters, one finds
\begin{align}\label{eq:fullZ}
   & Z(q,x,y,\eta,r)=1 +\eta\chi^{short}_{(2, 1)} X_{R_A}(y,r)  + 
\eta^2\chi_{(4, 0)} X_+ + \eta^2\chi_{(4, 1)}X_-  +\eta^2\chi_{(4, 2)}X_+ \nn  \\
&+ \eta^2\chi_{(5,0)}X_+  + \eta^2\chi_{(5, 1)} X_- +\eta^2\chi_{(5, 2)}(X_++X_-)  + \eta^2\chi_{(5, 3)} X_- + \eta^2\chi_{(6, 0)}X_+ \nn \\
    &+\eta^3\chi_{(6, 0)}Y_-  + \eta^2\chi_{(6, 1)}X_- +\eta^3\chi_{(6, 1)} (Y_++Z)+ 2\eta^2\chi_{(6, 2)}X_+  +\eta^3\chi_{(6, 2)} Z \nn \\
&  +\eta^2\chi_{(6, 3)}(X_++X_-)+\eta^3\chi_{(6, 3)}Y_+ + \eta^2\chi_{(6, 4)} X_+  + \eta^2\chi_{(7, 0)} X_+\nn \\
    &+\eta^3\chi_{(7, 0)}  (Y_-+Z) + \eta^2\chi_{(7, 1)}X_- +\eta^3\chi_{(7, 1)}(3Z+2Y_++Y_-)  + \eta^2\chi_{(7, 2)}(X_++X_-)\nn \\
  &+\eta^3\chi_{(7, 2)}(3Z+Y_++2Y_-)  + 2\eta^2\chi_{(7, 3)}(q, x)X_- +\eta^3\chi_{(7, 3)} (2Z+Y_++Y_-)\nn \\ 
  &  + \eta^2\chi_{(7, 4)}(X_++X_-)+\eta^3\chi_{(7, 4)} Z + \eta^2\chi_{(7, 5)}X_- + O(q^8)\,,
\end{align}
where we have defined the group character combinations
\begin{align}\label{eq:groupComb}
X_\pm & \equiv \frac{X_{R_A}^2(y,r)\pm X_{R_A}(y^2,r)}{2} \,, \\
Y_\pm & \equiv \frac{  X_{R_A}^3(y,r) \pm  3 X_{R_A}(y,r) X_{R_A}(y^2,r) +  2 X_{R_A}(y^3,r)}6 \,, \qquad 
Z\equiv \frac{ X_{R_A}^3(y,r) - X_{R_A}(y^3,r)}3 \,.\nn 
\end{align}
The $SU(r+1)$ representations under which the operators in \eqref{eq:fullZ} transform are encoded in the combinations \eqref{eq:groupComb}. 
For simplicity, we work out here the character decomposition for the primaries with $\Delta\leq 5$, which involve only the combinations $X_\pm$.
We first consider the general $SU(r+1)$ case with $r>3$. The cases $r=1,2,3$ are special and will be treated after. The decomposition of two adjoint representations reads
\begin{equation}\label{eq:rapp}
R_A\otimes R_A = R_+\oplus R_-\,,\quad R_+=\mathbf{1}\oplus R_A\oplus R_1\oplus R_3\,,\quad R_-= R_2 \oplus \overline{R}_2 \oplus R_A\,,
\end{equation}
where $\mathbf{1}$ is the singlet and $R_i$ are representations with Dynkin labels
\begin{align}
    R_1 & = (2,0,\ldots ,0,2) \,, \qquad\qquad\ \ \;  \text{dim}\; R_1  = \frac{(n_c+3)n_c^2(n_c-1)}{4}  \,, \nonumber\\
    R_2 & = (0,1,0,\ldots ,0,2) \,,  \qquad\qquad  \text{dim}\; R_2  = \frac{(n_c^2-4)(n_c^2-1)}{4}  \,,\nonumber \\
  \overline{R}_2 & = (2,0,\ldots, 0,1,0) \,, \qquad\qquad  \text{dim}\; \overline{R}_2  =  \text{dim}\; R_2\,, \\
     R_3 & = (0,1,0,\ldots ,0,1,0) \,, \qquad\quad  \text{dim}\; R_3 = \frac{n_c^2(n_c-3)(n_c+1)}{4}  \nonumber \\
 R_A & = (1,0,\ldots ,0,1) \,, \nonumber  \qquad\qquad \;\;\,  \text{dim}\; R_A  = n_c^2-1\,,
        \end{align}
with $n_c=r+1$. One can check that the following character decomposition holds,
\begin{align}
X_+ & = 1+ X_{R_1}(y,r) + X_{R_3}(y,r) + X_{R_A}(y,r)  \,,  \nn \\
X_- & = X_{R_2}(y,r) + X_{\overline R_2}(y,r) + X_{R_A}(y,r) \label{eq:chaASym} \,.
\end{align}
For $r=3$ the decomposition \eqref{eq:rapp} and \eqref{eq:chaASym} applies, but the Dynkin labels of $R_3$ are modified,
\begin{equation}
    R_3 = (0,2,0)\,, \qquad \qquad  \text{dim}\; R_3 = 20\,, \qquad (r=3)\,.
\end{equation}
For $r=2$ the representation $R_3$ does not exist, and we have
\begin{align}
    R_1 = (2,2)\,, \qquad \qquad \qquad\ \,  \text{dim}\; R_1 = 27\,, \qquad (r=2)\,, \nonumber \\
     R_2 = (3,0)\,, \quad  \overline{R}_2 = (0,3)\,, \qquad  \text{dim}\; R_2 =  \text{dim}\; \overline{R}_2=10\,, \qquad (r=2)\,.
\end{align}
For $r=1$ the decomposition trivializes,
\be
X_+  = 1+ X_{2}(y,r) \,, \qquad  X_-   = X_{1}(y,r) \,,\qquad \qquad (r=1) \,,
\ee
where the subscripts refer to the spin $j$ of the representation ($j=1$ is the adjoint).
We see that in the $SU(2)$ case the adjoint no longer appears in $X_+$. The character decomposition of $Y_\pm$ and $Z$ for $r=1$ is as follows:
$Y_+ = X_{1}(y,r)+X_{3}(y,r)$, $Y_-=1$, $Z= X_{1}(y,r)+X_{2}(y,r)$.
Hence the lightest adjoint scalar appears in the fifth row of \eqref{eq:fullZ}, it has $\Delta = 7$ and is given by the product of three currents and one derivative.

\section{Bulk-to-bulk gauge propagator}
\label{app:prop}

\subsection{Map to the notation of \cite{Allen1986}}
\label{app:mapAJ}

It is convenient to express propagators and their derivatives with the notation presented in \cite{Allen1986} for maximally symmetric spaces. 
Let us denote by $\mu(x,y)$ the geodesic distance, which can be expressed in terms of the chordal distance as 
\begin{equation}
\mu(x,y)=\cosh^{-1}\Big(1+u(x,y)\Big)\,.
\end{equation}
In this notation,  the building blocks are the parallel propagator $g^\nu_{\ \nu'}(x,y)$ transporting vectors along geodesics from $x$ to $y$, and the unit vectors $n_\nu(x,y)$ and $n_{\nu'}(x,y)$,   tangent to the geodesic at $x$ and $y$, respectively,
\begin{equation}
n_{\nu}\left(x, y\right)=\nabla_{\nu} \mu(x, y) \quad \text { and } \quad n_{\nu^{\prime}}\left(x, y\right)=\nabla_{\nu^{\prime}} \mu\left(x, y\right)\,.
\end{equation} 
 Any bitensor in a maximally symmetric space can be expressed as sums and products of these building blocks, with coefficients that are only functions of $\mu$, or equivalently of $u$. For example, a bitensor with an index in $x$ and an index in $y$, such as the gauge propagator $\Pi$, can be decomposed as 
 \begin{equation}
     \Pi_{\mu \mu'}(x,y)=\pi_0(u)g_{\mu \mu'}+\pi_1(u) n_\mu n_{\mu'}\,.
     \label{eq:bitensor}
 \end{equation}
In AdS space we also have \cite{Allen1986}
\begin{equation}
\begin{aligned}
\nabla_{\nu} n_{\mu}  =\frac{(1+u)}{\sqrt{u(u+2)}}\left(g_{\nu \mu}\right.&\left.\!-\,n_{\nu} n_{\mu}\right)\,, \qquad
\nabla_{\nu} n_{\mu^{\prime}}  =-\frac{1}{\sqrt{u(u+2)}}\left(g_{\nu \mu^{\prime}}+n_{\nu} n_{\mu'}\right)\,, \\
\nabla_{\nu} g_{\mu \mu'} & =-\frac{u}{\sqrt{u(u+2)}}\left(g_{\nu \mu} n_{\mu'}+g_{\nu \mu'} n_{\mu}\right),
\end{aligned}
\label{eq:useful}
\end{equation}
which are useful relations to compute derivatives of propagators. We can map between the parametrization \eqref{eq:propintr}  and the one given in \eqref{eq:bitensor} by the relations 
\begin{equation}
    \nabla_\mu u=\sqrt{u(u+2)} n_\mu \,,\qquad \nabla_{\mu'}\nabla_\mu u=-g_{\mu \mu'}+u\ n_\mu n_{\mu'}\,.
\end{equation}
We get
\begin{equation}\label{eq:mappPropComp}
\pi_0(u) = g_0(u)\,, \qquad \qquad \pi_1(u) = u(u+2) g_1(u) - u g_0(u)\,.
\end{equation}

\subsection{Spectral representation} 
\label{app:sprep}

We briefly review here the minimal properties of spin $\ell$ harmonic functions on AdS $\Omega_\nu^{(\ell)}$ needed for the derivation of the bulk gauge propagator, referring to \cite{Costa:2014kfa} for further details. We focus on $\ell=0,1$, which are the only cases of interest for us.
Harmonic functions on AdS can be conveniently defined in embedding space as suitable integrals over the boundary of two bulk-to-boundary propagators.  
They satisfy the relations
\be\label{eq:OmegaProp}
\begin{aligned}
 -\nabla_X^2 \Omega_{\nu AB}^{(1)}(X, Y)&=\left(\nu^2+\frac{d^2}{4}+1\right) \Omega_{\nu AB}^{(1)}(X, Y)\,, \\
 -\nabla_X^A \Omega_{\nu AB}^{(1)}(X, Y)&=0\,, \\
 -\nabla_X^2 \Omega_{\nu}^{(0)}(X, Y)&=\left(\nu^2+\frac{d^2}{4}\right) \Omega_{\nu}^{(0)}(X, Y)\,, 
\end{aligned}
\ee
as well as nice orthogonality properties. 
The harmonic functions can be written as
\be 
\Omega_\nu^{(\ell)}\left(X_1, X_2 ; W_1, W_2\right)=\frac{i \nu}{2 \pi}\left(G_{\frac{d}{2}+i \nu,\ell}\left(X_1, X_2 ; W_1, W_2\right)-G_{\frac{d}{2}-i \nu,\ell}\left(X_1, X_2 ; W_1, W_2\right)\right)\,,
\label{eq:harmonic}
\ee
where $G_{\Delta,\ell}$ is the analytic continuation for complex $\Delta$ of the bulk propagators
for a massive spin $\ell$ field \cite{Costa:2014kfa}.

Let us review how to get the $N$ and $D$ bulk propagators for a scalar field with mass $m^2 = \Delta (\Delta-d)$ using the spectral representation. We take $\Delta>d/2$.
In embedding space the equation of motion reads
\begin{equation}\label{eq:eomScalar}
    (-\nabla^2 + m^2)\Pi(X_1,X_2) = \delta(X_1,X_2)\,.
\end{equation}
We look for a particular solution of \eqref{eq:eomScalar} by writing
\begin{equation}\label{eq:srScalar}
\Pi(X_1,X_2)=\int_{-\infty}^{\infty}\! d\nu\, a_0(\nu) \Omega_\nu^{(0)}(X_1,X_2)\,,
\end{equation}
and the spectral representation of the delta function
\begin{equation}
\delta(X_1,X_2) =\int_{-\infty}^{\infty}\! d\nu\, \Omega_\nu^{(0)}(X_1,X_2)\,.
\end{equation}
Using the third relation in \eqref{eq:OmegaProp} it is immediate to determine $a_0$:
\begin{equation}
a_0(\nu) =\frac{1}{\nu^2+\Big(\Delta - \frac d2\Big)^2}\,.
\end{equation}
The integral in $\nu$ in \eqref{eq:srScalar} can be performed using \eqref{eq:harmonic} and residue theorem. Given the behavior of $G_{\Delta,0}$ as $\Delta$ goes to infinity in the complex plane,
we have to close the contour in the lower and upper half-plane for $G_{d/2+i \nu,0}$ and $G_{d/2-i \nu,0}$ respectively. 
The only poles are the ones given by $a_0(\nu)$, at $\nu_0 = i (\Delta-d/2)$ and $\nu_0^*$, the two residues giving the same contribution. We get
\be\label{eq:scalarD}
\Pi^{({ D})}(X_1,X_2) =2 \int_{C_{\nu_0^*}} \!\!\!d\nu \, a_0(\nu) \frac{i \nu}{2\pi}  G_{\frac d2+i \nu,0}(X_1,X_2) = G_{\frac d2+i \nu_0^*,0}(X_1,X_2) =G_{\Delta,0}(X_1,X_2)\,,
\ee
where $C_{\nu_0^*}$ is a small circle around $\nu_0^*$. The function $\Pi^{({ D})}$ is identified as the bulk propagator with $D$ bc. The $N$ bulk propagator $\Pi^{( N)}$ is determined by noticing that $\Omega_{\nu_0}^{(0)}$ is a solution of the homogeneous equation of motion, so $\Pi(X_1,X_2)+ c \, \Omega_{\nu_0}^{(0)}$ is a solution of \eqref{eq:srScalar} for any constant $c$. Demanding that $\phi\sim u^{\Delta-d}$ as $u\rightarrow\infty$ fixes the constant to be $c = 2\pi/(i \nu_0)$. This can also be written as
\begin{align}\label{eq:scalarN}
\Pi^{({ N})}(X_1,X_2) & = \Pi^{({D})}(X_1,X_2) -2  \int_{C_{\nu_0}}\!\! \!d\nu \, a_0(\nu)  \frac{i \nu}{2\pi} ( G_{\frac d2+i \nu,0}(X_1,X_2) -  G_{\frac d2-i \nu,0}(X_1,X_2) ) \\
& = 2 \int_{C_{\nu_0}} \!\!\!d\nu \, a_0(\nu)\frac{i \nu}{2\pi}  G_{\frac d2+i \nu,0}(X_1,X_2) = G_{\frac d2+i \nu_0,0}(X_1,X_2) =G_{d-\Delta,0}(X_1,X_2)\,. \nn
\end{align}
We see that the $N$ bulk propagator can be expressed, in the spectral representation, by the same integrand of the $D$ bulk propagator, but evaluated at the ``opposite" residues.
In this way, the correct boundary behavior is obtained. The same mechanism works for massive spin 1 and massless gauge propagators. 
In particular, the massive spin 1 propagator can be decomposed as follows in terms of harmonic functions \cite{Costa:2014kfa},
\begin{align}\label{eq:Pi1Proca}
\Pi_{\Delta,1}(X_1,X_2;W_1,W_2)&=\int\! d\nu\, \frac{1}{\nu^2+(\Delta-d/2)^2}\Omega_\nu^{(1)}(X_1,X_2;W_1,W_2)\\
& +(W_1\cdot \nabla_1)(W_2\cdot \nabla_2)\int\! d\nu\,\frac{1}{(\Delta-1)(\Delta-d+1)}\frac{1}{\nu^2+d^2/4}\Omega_\nu^{(0)}(X_1,X_2)\,. \nn
\end{align}
As for the scalar case above, $N$ and $D$ propagators are obtained by appropriately choosing the contour of the $\nu$ integration in the two cases.

\subsection{General expression in any $\xi$-gauge} 
\label{app:anygauge}

We report below the expression for the functions $g_0^{{(D,N)}}(u)$ and $g_1^{{( D,N)}}(u)$ entering the bulk-to-bulk gauge propagators \eqref{eq:Pig01Def} and \eqref{eq:propintr} for any $\xi$-gauge. For clarity, we split them in their transverse ($\xi$-independent) and longitudinal (proportional to $\xi$) components,
\be
g_i(u)=g_{i,\perp}(u)+\xi\,  g_{i,L}(u)\,, \qquad i=0,1\,.
\ee
The Dirichlet bulk-to-bulk propagator is given by 
\begin{equation}\label{eq:bulkDanyxi}
  \begin{aligned}
    g_{0,\perp}^{{( D)}} & =\frac{ \Gamma \left(\frac{d+1}{2}\right) \left(-\frac{1}{d}-\psi \left(\frac{d}{2}\right)+\psi (d)+u (u+2)-\frac{1}{2} \log (4 u (u+2))\right)}{2\pi ^{\frac{d+1}{2}}(d-2) (u (u+2))^{\frac{d+1}{2}}}\\
    &\quad +\frac{\Gamma \left(\frac{d+1}{2}\right) \left(\frac\partial{\partial b}+2\frac\partial{\partial c}\right)\left.
   {}_2F_1\left(\frac{d+1}{2},\frac{d+2}{2}+b,\frac{d+2}{2}+c,\frac{1}{(u+1)^2}\right)\right|_{b=c=0}}{4 \pi ^{\frac{d+1}{2} }  (d-2)(u+1)^{d+1}}\,,\\
   g_{0,L}^{{( D)}}&=\frac{  \Gamma \left(\frac{d+1}{2}\right) \left(H_{\frac{d}{2}}-\frac{1}{d}-\psi (d)+\log (2 (u+2))-\gamma_E \right)}{2\pi ^{\frac{d+1}{2}} d (u(u+2))^{\frac{d+1}{2}}}\\
   &\quad -\frac{  \Gamma \left(\frac{d+1}{2}\right) \left(\frac\partial{\partial a}+2\frac\partial{\partial c}\right)\left.
   {}_2F_1\left(d+1+a,\frac{d+1}{2},d+1+c,-\frac{2}{u}\right)\right|_{a=c=0}}{2\pi ^{\frac{d+1}{2}} du^{d+1}}\,,\\
   g_{1,\perp}^{{( D)}}&=\frac{ (u+1)  \Gamma \left(\frac{d+1}{2}\right) \left(  u (u+2)-\frac12 \log \left(\frac{u (u+2)}{(u+1)^2}\right)-  (d+1) \left(\psi \left(\frac{d}{2}\right)-\psi (d)+ \log (2 (u+1))\right)-\frac1d\right)}{2\pi ^{\frac{d+1}{2}} (d-2)  (u (u+2))^{\frac{d+3}{2}}}\\
   &\quad +\frac{  \Gamma \left(\frac{d+1}{2}\right) \left(\frac\partial{\partial b}+2\frac\partial{\partial c}\right)\left.
   {}_2F_1\left(\frac{d+1}{2},\frac{d+2}{2}+b,\frac{d+2}{2}+c,\frac{1}{(u+1)^2}\right)\right|_{b=c=0}}{4\pi ^{\frac{d+1}{2}} (d-2) u (u+2)(u+1)^{d}}\\
   &\quad +\frac{d \Gamma \left(\frac{d+1}{2}\right) \left(\frac\partial{\partial a}+\frac\partial{\partial b}+2\frac\partial{\partial c}\right)\left.
   {}_2F_1\left(\frac{d+1}{2}+a,\frac{d}{2}+b,\frac{d}{2}+c,\frac{1}{
   (u+1)^2}\right)\right|_{a=b=c=0}}{4\pi ^{\frac{d+1}{2}} (d-2) u (u+2) (u+1)^{d} }\,,\\
   g_{1,L}^{{( D)}}&=-\frac{\partial}{\partial u}g_{0,L}^{{( D)}}\,.
    \end{aligned}
\end{equation}
We have checked that the bulk propagator in the Feynman gauge $\xi=1$ agrees with what is found in \cite{Allen1986}.

The Neumann bulk-to-bulk propagator is given by 
\be\label{eq:bulkNanyxi}
\begin{aligned}
g_{0,\perp}^{{(N)}} & = \frac{ {}_2F_1\left(\frac{1}{2},1;1-\frac{d}{2};\frac{1}{(u+1)^2}\right) \left(-H_{1-\frac{d}{2}}-\frac{2}{d-1}+\frac{1}{d}-\log (2 (u+1))+1\right)}{4 \pi ^{d/2} \Gamma \left(2-\frac{d}{2}\right) (u+1)}\\
&\quad -\frac{(d-2)  u (u+2) \, _2\tilde{F}_1\left(\frac{3}{2},2;2-\frac{d}{2};\frac{1}{(u+1)^2}\right)}{2 \pi ^{d/2}(d-1) d (u+1)^3}\\
&\quad +\frac{(d-1) \pi ^{-d/2} \left(\frac{\partial}{\partial a}+\frac{\partial}{\partial b}+2\frac{\partial}{\partial c}\right)\left.{}_2F_1\left(\frac{1}{2}+a,1+b,2-\frac{d}{2}+c,\frac{1}{(u+1)^2}\right)\right|_{a=b=c=0}}{8
   (d-2) \Gamma \left(2-\frac{d}{2}\right)(u+1) }\\
   &\quad +\frac{ \left(\frac{\partial}{\partial a}+\frac{\partial}{\partial b}+2\frac{\partial}{\partial c}\right)\left.{}_2F_1\left(\frac{3}{2}+a,1+b,2-\frac{d}{2}+c,\frac{1}{(u+1)^2}\right)\right|_{a=b=c=0}}{4\pi ^{d/2}
   (d-2)^2 (u+1) \Gamma \left(1-\frac{d}{2}\right)}\,, \\
g_{0,L}^{{(N)}} & = \frac{  \Gamma \left(\frac{d+1}{2}\right) \left(H_{-\frac{d}{2}-\frac{1}{2}}+H_{\frac{d}{2}}-\frac{2}{d}+\pi  \tan \left(\frac{\pi  d}{2}\right)-2 (\psi (d)+\gamma_E )+\log \left(\frac{4 (u+2)}{u}\right)\right)}{2\pi ^{\frac{d+1}{2}} d  (u(u+2))^{\frac{d+1}{2}}}\\
&\quad +\frac{ \, _2F_1\left(1,\frac{1-d}{2};1-d;-\frac{2}{u}\right) \left(\pi  \cot \left(\frac{\pi  d}{2}\right)+2 \psi (d)-\psi \left(\frac{d+1}{2}\right)+\log \left(\frac{u}{2}\right)+\gamma_E \right)}{\pi ^{d/2} d^2 u \Gamma \left(-\frac{d}{2}\right)}\\
&\quad -\frac{ \left(\frac{\partial}{\partial a}+\frac{\partial}{\partial b}+2\frac{\partial}{\partial c}\right)\left.{}_2F_1\left(1+a,\frac{1-d}{2}+b,1-d+c,-\frac{2}{u}\right)\right|_{a=b=c=0}}{\pi ^{d/2}d^2 u \Gamma
   \left(-\frac{d}{2}\right)}\\
   &\quad +\frac{  \Gamma \left(\frac{d+1}{2}\right) \frac\partial{\partial b}\left.{}_2F_1\left(d+1,\frac{d+1}{2}+b,d+1,-\frac{2}{u}\right)\right|_{b=0}}{2\pi ^{\frac{d+1}{2}} d u^{d+1}}\,,\\
g_{1,\perp}^{{ (N)}} & = \frac{\left(d^2-1\right)  \, _2F_1\left(\frac{1}{2},1;2-\frac{d}{2};\frac{1}{(u+1)^2}\right) \left(H_{1-\frac{d}{2}}-\frac{u (u+2)}{d^2-1}+\log (2 (u+1))-1\right)}{4 \pi ^{d/2}(d-2) u^2 (u+2)^2 \Gamma \left(2-\frac{d}{2}\right)}\\
&\quad -\frac{ \left(d+(u+1)^2\right) \left(H_{1-\frac{d}{2}}+\log (2 (u+1))-1\right)}{4\pi ^{d/2} u^2 (u+2)^2 \Gamma \left(2-\frac{d}{2}\right)}\\
&\quad +\frac{(d-1) \left(\frac{\partial}{\partial a}+\frac{\partial}{\partial b}+2\frac{\partial}{\partial c}\right)\left.{}_2F_1\left(\frac{1}{2}+a,1+b,2-\frac{d}{2}+c,\frac{1}{(u+1)^2}\right)\right|_{a=b=c=0}}{8 \pi ^{d/2} 
   (d-2) u (u+2) \Gamma \left(2-\frac{d}{2}\right)}\\
   &\quad -\frac{ \left(d+(u+1)^2\right) \left(\frac{\partial}{\partial a}+\frac{\partial}{\partial b}+2\frac{\partial}{\partial c}\right)\left.{}_2F_1\left(\frac{3}{2}+a,1+b,2-\frac{d}{2}+c,\frac{1}{(u+1)^2}\right)\right|_{a=b=c=0}}{8\pi ^{d/2}
   (d-2) u (u+1)^2 (u+2) \Gamma \left(2-\frac{d}{2}\right)}\,, \\
g_{1,L}^{ (N)} & =-\frac{\partial}{\partial u} g_{0,L}^{(N)}\,. 
\end{aligned}
\ee

\section{Boundary RG flow from bulk dynamics}\label{app:boundaryRG}

In this appendix, we show how an RG flow can be induced in the boundary CFT when an irrelevant boundary operator becomes marginal.
The analysis will be given for a general bulk and boundary theory, with the assumption that in the UV the bulk theory is conformal and the boundary theory has 
no relevant deformations. The latter assumption is not essential, but it simplifies the analysis that follows.
We denote by $\hat O$ the lowest dimensional irrelevant scalar singlet operator of the boundary CFT and by 
$O$ the leading bulk operator, whose coupling $\widetilde \lambda$ govern the CFT data of the boundary theory. We take $\lambda = \widetilde \lambda\, L^{d+1-\Delta_O}$, where $\Delta_O$ is the scaling dimension of $O$ in the bulk CFT at $\tilde{\lambda}= 0$, in such a way that $\lambda$ is dimensionless.

Suppose then that there exists a value $\lambda_{\rm crit}$ (or alternatively a critical AdS length $L_{\rm crit}$), where $\Delta_{\hat O}(\lambda_{\rm crit})=d$.
Let us denote by $\eta\ll 1$ the coupling associated to $\hat O$ when this is close to marginal, and by $\delta \lambda = \lambda-\lambda_{\rm crit}$ the deviation of the
coupling from its critical value. We take
\be
\delta \lambda \ll \eta \ll 1\,, \qquad \qquad \delta \lambda \sim \eta^2\,,
\ee
and use $\eta$ as expansion parameter. We determine the beta function $\beta_\eta$ of the coupling $\eta$ up to order $\eta^2$ by using techniques similar to those employed in conformal perturbation theory, i.e. we expand a correlation function of bulk operators around the bulk critical theory in absence of boundary deformations. 
A simple choice is to consider the one-point function of the bulk operator $O$ itself, the one associated with the deformation $\delta \lambda$. We have
\begin{align}\label{eq:betay1}
\langle O(x_1) \rangle_{\delta \lambda,\eta} & = \langle O(x_1) \rangle_{0} - \eta  \langle O(x_1) \int\! d\hat x \, \hat O( x) \rangle_{0}
+\frac{\eta^2}2  \langle O(x_1) \int\! d\hat x \, \hat O( x)  \int\! d\hat y \, \hat O( y) \rangle_{0} \nn \\
& \quad - \delta \widetilde \lambda  \langle O(x_1) \int\! dx \, O(x) \rangle_{0} + \ldots\,,
\end{align}
where the subscript $0$ means that the correlator is evaluated at $\delta \widetilde \lambda = \eta = 0$ and $d\hat x \equiv d^d x$ denotes the measure at the boundary. The renormalization of the boundary coupling $\eta$ is determined by the short-distance behavior of the above correlators, which is fixed using the OPE and the bOPE expansions. The latter still exists despite the bulk theory is generally non-conformal at $\lambda=\lambda_{\text{crit}}$.
The contribution to $\beta_\eta$ coming from the third term in the first row of \eqref{eq:betay1} is computed using standard techniques of conformal perturbation theory (see e.g. chapter 5 
of \cite{Cardy:318508}). Short-distance divergences occur when $x$ approaches $y$. We can then use the OPE to rewrite that term as 
\be\label{eq:betay1a}
 \int\! d\hat x \, \hat O(x)  \int\! d\hat y \, \hat O( y) \approx \int\! d\hat x \, \int_{| w|\geq a} \! \!\! d\hat w\, \Big(C_{\hat O\hat O}  w^{-2d} + C_{\hat O\hat O}^{\quad \hat O}   w^{-d}\hat O( x) + \ldots \Big)  \,,
\ee
where $ w_i = ( x- y)_i$, $a$ is a short-distance cut-off, $C_{\hat O\hat O}$ and $C_{\hat O\hat O}^{\quad \hat O}$ are the coefficients entering the two-point function and the OPE coefficient of the three-point function of 
$\hat O$.\footnote{Note that we are not assuming unit-normalized two-point functions, so the OPE coefficient $C_{\hat O\hat O\hat O}$ entering a three-point function and the one coming from the OPE $C_{\hat O\hat O}^{\quad \hat O}$ are not identical. One has
\be
C_{\hat O\hat O}^{\quad \hat O} = \frac{C_{\hat O\hat O\hat O}}{C_{\hat O\hat O}}\,.
\ee
A similar relation occurs between $B_{O\hat O}$ entering a two-point bulk-to-boundary two-point function and the coefficient $B_O^{\; \hat O}$ coming from the bOPE.} 
Universal contributions to $\beta_\eta$ arise from the second term in \eqref{eq:betay1a}.\footnote{For example, using the regularization \eqref{eq:maintool}, the first term in eq.~\eqref{eq:betay1a} is UV finite.} We have
\be\label{eq:betay2}
 \int\! d\hat x \, \hat O(x)  \int\! d\hat y \, \hat O(y)\Big|_{{\rm div}} \approx -\Omega_{d-1} C_{\hat O\hat O}^{\quad \hat O} \log(a)  \int\! d\hat x \,   \hat O(x)  \,.
\ee
The first term in the second row of \eqref{eq:betay1} is UV divergent when the bulk operator approaches the boundary. In this limit we can expand
the bulk field in terms of boundary operators using the bOPE:
\be\label{eq:betay3}
L^{\Delta_O} O(x) =  \sum_{\hat O_k} z^{\Delta_{\hat O_k}} B_O^{\hat O_k} \hat O_k(x) =  B_O^{\; \hat{\mathbf{1}}} +B_O^{\hat O} z^d \hat O(x)+\ldots  \,.
\ee
We then have
\be\label{eq:betay3pt5}
O(x_1) \int\! dx \, O(x) \approx L^{d+1-\Delta_O} O(x_1) \int\! d\hat x \int_a^\infty \!\frac{dz}{z^{d+1}} \, \Big( B_O^{\; \hat{\mathbf{1}}} +B_O^{\; \hat O} z^d \hat O( x)+\ldots \Big)\,.
\ee
As in \eqref{eq:betay1a}, universal logarithmic contributions arises only from the second term in \eqref{eq:betay3}, which gives
\be\label{eq:betay4}
O(x_1) \int\! dx \, O(x)\Big|_{{\rm div}} \approx - \log(a)  L^{d+1-\Delta_O}  B_O^{\; \hat O} \, O(x_1) \int\! d\hat x \, \hat O(x)\,.
\ee
In general both the 3-point OPE and the bOPE coefficients $C_{\hat O \hat O \hat O}$ and $B_{O \hat O}$ depend on $\lambda$.
From \eqref{eq:betay1}, \eqref{eq:betay2} and \eqref{eq:betay4} we immediately get  
\be
\beta_\eta = - a\frac{d\eta}{da} = c_1 \eta^2 + c_2 \delta \lambda \,,
\ee
where
\be\begin{split}
c_1 & = \frac 12 \Omega_{d-1} C_{\hat O\hat O}^{\quad \hat O}(\lambda_{\rm crit}) \,, \\
c_2 & = - B_O^{\; \hat O}(\lambda_{\rm crit}) \,.
\end{split}
\ee

\section{Tools for the computation of Witten diagrams}

Developing tools for the computation of Witten diagrams at loop level is an active field of research in itself, see e.g. \cite{Penedones:2010ue,Aharony:2016dwx,Yuan:2017vgp,Yuan:2018qva,Giombi:2017hpr,Cardona:2017tsw,Liu:2018jhs,Bertan:2018khc,Bertan:2018afl,Ghosh:2019lsx,Ponomarev:2019ofr,Carmi:2019ocp,Albayrak:2020bso,Herderschee:2021jbi,Carmi:2021dsn,Heckelbacher:2022fbx,Carmi:2024tzj,Cacciatori:2024zbe}. In this appendix, we provide details on several technical points required to compute the Witten diagrams presented in sections \ref{sec:gammaJJ} and \ref{sec:cj}. 

\subsection{Mass shift diagram}
\label{app:massshift}
We want to compute the integral corresponding, modulo some prefactors that we will make explicit later, to  the mass shift of a scalar field of dimension $\Delta$, which is 
\be\label{eq:IntIDelta}
\mathcal{I}_\Delta=\int d X \frac{1}{\left(-2 P_1 \cdot X\right)^\Delta} \frac{1}{\left(- 2 P_2 \cdot X\right)^\Delta}\,,
\ee
where $X$ is a point in the bulk and $P_1$, $P_2$ are points at the boundary. 
The integral \eqref{eq:IntIDelta} is divergent, but it can be regulated for $\Delta\neq d/2$ by 
putting $P_1$ and $P_2$ at distance $z_{1,2}=\delta\ll 1$ from the boundary.
Consider then the analog bulk integral
\be
\widetilde{\mathcal{I}}_\Delta=\int \! d X \,  G_{\Delta,0}(X_1,X ) G_{\Delta,0}\left(X, X_2\right)\,,
\label{eq:massshif}
\ee
with $X_1$ and $X_2$ points in the bulk and $G_{\Delta,0}$ the bulk-to-bulk propagator of a scalar field with dimension $\Delta$. 
We assume from now that $\Delta \neq d/2$. In spectral representation, the scalar propagator 
with $D$ bc can be written as
\begin{equation}\label{eq:specG0}
     G_{\Delta,0}\left(X_1, X_2\right)=\int_{-\infty}^{+\infty} \! d \nu \, \frac{1}{\nu^2+\left(\Delta-\frac{d}{2}\right)^2} \Omega_\nu^{(0)}\left(X_1, X_2\right)\,.
\end{equation}
Using \eqref{eq:specG0} and the orthogonality relation for harmonic functions
\begin{equation}
    \begin{aligned}
        \int \! dX\,\Omega_\nu^{(0)}(X_1,X)\Omega_{\nu'}^{(0)}(X,X_2)=\frac{\delta(\nu -\nu')+\delta(\nu +\nu')}{2}\Omega_\nu^{(0)}(X_1,X_2)\,,
    \end{aligned}
\end{equation}
we can rewrite the integral \eqref{eq:massshif} as
\begin{equation}
\begin{aligned}
\widetilde{\mathcal{I}}_\Delta&=\int_{-\infty}^{+\infty} d \nu \frac{1}{\left(\nu^2+\left(\Delta-\frac{d}{2}\right)^2\right)^2} \Omega_\nu^{(0)}\left(X_1, X_2\right) 
=-\frac{1}{2 \Delta-d} \frac{d}{d \Delta} G_{\Delta,0}\left(X_1, X_2\right) \,.
\end{aligned}
\end{equation}
Taking the external points to the boundary we get
\begin{equation}
\mathcal{I}_{\Delta}=\lim _{z_{1,2} \rightarrow \delta}\left[\left(z_1 z_2\right)^{-\Delta}\widetilde{\mathcal{I}}_\Delta\right] =-\frac{1}{2 \Delta-d} \frac{1}{(C_0(\Delta))^2} \lim _{z_{1,2} \rightarrow \delta}\left[\left(z_1 z_2\right)^{-\Delta} \frac{d}{d \Delta} G_{\Delta,0}\left(X_1, X_2\right)\right]\,.
\label{eq:Idelta}
\end{equation}
We now use the relation between the bulk-to-bulk and the bulk-to-boundary propagator
\begin{equation}
 G_{\Delta,0}\left(X_i, X\right) \underset{z_i \rightarrow \delta}{\sim}  z_i^\Delta\frac{C_0(\Delta)}{\left(-2 P_i \cdot X\right)^{\Delta}}  \,,
 \label{eq:scalarbulktobound}
\end{equation}
where 
\be C_0(\Delta) = \frac{\Gamma(\Delta)}{2\pi^{\frac d2}\Gamma\Big(1-\frac d2 + \Delta\Big)}\,.
\ee
We compute the limit by using \eqref{eq:scalarbulktobound}, taking both points to the boundary,
\begin{equation}
 G_{\Delta,0}\left(X_1, X_2\right) \underset{z_{1,2} \rightarrow 0}{\sim}  z_1^\Delta z_2^\Delta\frac{C_0(\Delta)}{\left(-2 P_1 \cdot P_2\right)^{\Delta}}  \,.
\end{equation}
This gives the final result
\begin{equation}
\mathcal{I}_{\Delta}=\frac{1}{d-2 \Delta} \frac{1}{(C_0(\Delta))^2}\left[-C_0(\Delta) \log\Big( \frac{-2 P_1 \cdot P_2}{\delta^2}\Big)+\frac{d}{d \Delta} C_0(\Delta)\right] \frac{1}{\left(-2 P_1  \cdot P_2\right)^{\Delta}} \,, 
\end{equation}
which equals \eqref{eq:massshiftresult} in the main text.

\subsection{Integration by parts in AdS space}
\label{app:intparts}
We show that integration by parts in AdS gives vanishing boundary terms if the derivatives that we are moving act on bulk-to-bulk propagators. This is not the case when bulk-to-boundary propagators are involved in the diagrams. Let us consider the following integral,
\begin{equation}
\intads dx\ \langle A_{i}^{a_1}(x_1) \nabla_\nu A_\lambda^a(x)\rangle A^{\nu b} A^{\lambda c}(x)\,.
\end{equation}
Let us integrate by parts and focus only on the boundary term, which is
\begin{equation}
\lim_{\delta\rightarrow 0}\int_{z=\delta}d\hat{x}\frac{1}{\delta^{d+1}}\delta^2\left(-\langle A_{i_1}^{a_1}(x_1)  A_\lambda^a(\hat{x};\delta)\rangle A_z^b A_\lambda^c(\hat{x},\delta)\right)\,,
\end{equation}
where $d\hat x \equiv d^d x$ denotes the measure at the boundary.
Now, we have
\begin{equation}
A_i^a(x,\delta)\underset{\delta \to 0}{\sim} \delta^{d-2}J_i^a(x) \,,
\qquad
A_z^a(x,\delta)\underset{\delta \to 0}{\sim}\delta^{d-1}b^a(x) \,,
\label{eq:boundvec}
\end{equation}
where $J_i^a$ are the boundary currents and $b^a$ is some function which does not depend on $\delta$. This implies that the boundary contribution vanishes for sufficiently large $d$. We can repeat the same argument for the ghost vertex,
\begin{equation}
\intads dx\ \langle A_{i_1}^{a_1}(x_1)A_\nu^b(x)\rangle \nabla^\nu \bar{c}^a c^c\,.
\end{equation}
Let us integrate by parts and focus only on the boundary term, which is
\begin{equation}
\lim_{\delta\rightarrow 0}\int_{z=\delta}d\hat{x}\frac{1}{\delta^{d+1}}\delta^2\left(-\langle A_{i_1}^{a_1}(x_1) A_z^b({x},\delta)\rangle \bar{c}^a c^c({x},\delta)\right)\,.
\end{equation}
Now we have eq.\eqref{eq:boundvec} and
\begin{equation}
\bar{c}(x,\delta)\underset{\delta \to 0}{\sim} \delta^{d}\hat{\bar{c}}_d(x) \,,\quad c(x,\delta)\underset{\delta \to 0}{\sim} \delta^{d}\hat{c}_d(x)\,,
\end{equation}
with $c_0$ and $\bar{c}_0$ generic {Grassmann-odd} functions independent of $\delta$,
which again makes the boundary term vanish.

\subsection{A useful integral}
\label{app:usefulint}
Let us compute the following integral, which enters in  the triple diagram contribution to the $J^aJ^b$ two-point function,
\begin{equation}
\mathcal{K}^{(k)}=\int dX\ dY\ (-2P_1\cdot X)^{-\Delta_1} (-2P_1\cdot Y)^{-\Delta_3} (-2P_2\cdot X)^{-\Delta_2} (-2P_2\cdot Y)^{-\Delta_4} \zeta^{\Delta_k}\,,
\end{equation}
with $\zeta$ as defined in \eqref{eq:zetaDef}.
We exploit  AdS symmetries to simplify the expression of this integral. We begin by using translation symmetry to set $P_2=(1,0,0)$. We then use inversion, which acts in embedding coordinates by exchanging $X^+$ and $X^-$ coordinates. 
This corresponds to taking $P_2=(0,1,0)$ and implies that, given 
\begin{equation}
        X=\frac{1}{z}\left(1,x^2+z^2,x^i\right)\,,\qquad \qquad
        Y=\frac{1}{w}\left(1,y^2+w^2,y^i\right)\,,
\end{equation}
we have 
\begin{equation}
(-2P_2\cdot X)=\frac{1}{z}\,, \qquad \qquad (-2P_2\cdot Y)=\frac{1}{w}\,.
\label{eq:p2inv}
\end{equation}
The other scalar products are instead given by 
\begin{equation}(-2P_1\cdot X)=\frac{(x-y_1)^2+z^2}{zy_1^2}\,,\qquad(-2P_1\cdot Y)=\frac{(y-y_1)^2+w^2}{wy_1^2}\,.
\label{eq:p1inv}
\end{equation}
Note that the chordal distance and the metric determinant are not affected by this transformation,
\begin{equation}
u= \frac{(z-w)^2+(x-y)^2}{2 z w}\, \quad \longrightarrow  \quad \zeta=\frac{2 z w}{w^2+z^2+(x-y)^2}\,.
\end{equation}
Using the notation of eq.\eqref{eq:notDelta} in the main text, we can rewrite $\mathcal{K}^{(k)}$ as follows,
\begin{equation}
\begin{aligned}
    \mathcal{K}^{(k)}=\int \frac{dz d^d x}{z^{d+1}} \int \frac{dw d^d y}{w^{d+1}} \frac{2^{\Delta_k} y_1^{2\Di_{13}} w^{\Delta_{34k}}z^{\Di_{12k}}}{(z^2+(x-y_1)^2)^{\Di_1}(w^2+(y-y_1)^2)^{\Di_3}(w^2+z^2+(x-y)^2)^{\Di_k}} \,.
    \end{aligned}
    \end{equation}
We introduce Feynman parameters to rewrite the relevant terms in the denominator as
\begin{equation}
\begin{aligned}
\frac{\Gamma(\Di_{3k})}{\Gamma(\Di_3)\Gamma(\Di_k)}\int_0^1 d \alpha(1-\alpha)^{\Di_3-1}\alpha^{\Di_k-1}(w^2+y^2+\alpha(z^2+(1-\alpha)(x-y_1)^2)\,.
\end{aligned}
\end{equation}
The integral over $y$ and $w$ is now straightforward. The result, after rescaling $z\rightarrow \sqrt{1-\alpha} \ z$, is 
\begin{align}
    \mathcal{K}^{(k)}=\int \frac{dz d^d x}{z^{d+1}} \int_0^1 d\alpha  \frac{\Gamma\left(\frac{\Di_{3k,4}}{2} \right)\Gamma\left(\frac{\Di_{34k}-d}{2} \right)}{\Gamma(\Di_3)\Gamma(\Di_k)}\frac{2^{\Delta_k-1} \pi^\frac{d}{2}y_1^{2\Di_{13}} z^{\Di_{12k}}\alpha^{\frac{\Delta_{4k,3}-2}{2}}(1-\alpha)^\frac{\Di_{1234}-d-2}{2}}{((1-\alpha)z^2+(x-y_1)^2)^{\Di_1}(z^2+(x-y_1)^2)^{\frac{\Di_{3k,4}}{2}}}\,. \nn
    \end{align}
    Performing the integral over $\alpha$ gives
    \begin{equation}
    \begin{aligned}
        \mathcal{K}^{(k)}&=\frac{
    \Gamma\left(\frac{ \Di_{1234}-d }{2} \right)\Gamma\left(\frac{\Di_{3k,4}}{2}\right)
\Gamma\left(\frac{\Di_{4k,3}}{2}\right)
\Gamma\left(\frac{ \Di_{34k}-d}{2} \right)}{\Gamma(\Di_3) \Gamma(\Di_k)\Gamma\left(\frac{ \Di_{124k}+\Di_4-d}{2}\right)}\\
     &\qquad \int \frac{dz d^d x}{z^{d+1}}\frac{2^{\Di_k-1} \pi^\frac{d}{2} y_1^{ 2\Di_{13}}  z^{\Di_{12k}}\ _2F_1\left(\Di_1, 
    \frac{\Di_{4k,3}}{2}, 
    \frac{ \Di_{124k}+\Di_4-d}{2}, 
    \frac{z^2}{(x - y_1)^2 + 
     z^2}\right)}{((x - 
       y_1)^2 + z^2)^\frac{\Di_{13k,4}+\Di_1}{2}} \,.
\end{aligned}
    \end{equation}
Now we can replace ${}_2F_1$ by its series expansion,
\begin{equation}
_2F_1(a,b,c,z)=\sum_{m=0}^\infty\frac{(a)_m(b)_m}{ m!(c)_m}z^m\,,
\end{equation}
and go back to embedding coordinates using  eq.\eqref{eq:p1inv}. Imposing the condition $\Di_{13}= \Di_{24}$, which is satisfied by each of the integrals of interest, we finally get \eqref{eq:I4k} reported in the main text.

\bibliographystyle{JHEP}
\bibliography{GaugeTheoriesInAdS4}

\end{document}